\documentclass[pra,twocolumn,showpacs,letterpaper,showpacs,superscriptaddress]{revtex4}
\usepackage{amsmath,braket,amssymb,graphicx,stackrel,color,hyperref}

\definecolor{darkblue}{rgb}{0.,0.,0.5}
\definecolor{darkred}{rgb}{0.5,0.,0.}
\definecolor{darkgreen}{rgb}{0.,0.5,0.}

\hypersetup{pdftitle={CP-Plasmons},colorlinks=true,linkcolor=darkred,citecolor=darkblue,urlcolor=darkblue}

\newcommand{\PP}{\mathrm{P}}

\begin{document}

\title{Nonequilibrium Casimir-Polder plasmonic interactions}

\author{Nicola Bartolo}
\email{nicola.bartolo@univ-paris-diderot.fr}
\affiliation{Universit\'{e} Paris Diderot, Sorbonne Paris Cit\'{e}, Laboratoire Mat\'{e}riaux et Ph\'{e}nom\`{e}nes Quantiques, CNRS-UMR7162, 75013 Paris, France}

\author{Riccardo Messina}
\affiliation{Laboratoire Charles Coulomb, UMR 5221 Universit\'{e} de Montpellier and CNRS, F-34095 Montpellier, France}

\author{Diego A. R. Dalvit}
\affiliation{Theoretical Division, MS B213, Los Alamos National Laboratory, Los Alamos, New Mexico 87545, USA}

\author{Francesco Intravaia}
\affiliation{Max-Born-Institut, 12489 Berlin, Germany}

\begin{abstract}
We investigate how the combination of nonequilibrium effects and material properties impacts on the
Casimir-Polder interaction between an atom and a surface. By addressing systems with temperature inhomogeneities and laser interactions, we show that nonmonotonous energetic landscapes can be produced where barriers and minima appear. Our treatment provides a self-consistent quantum theoretical framework for investigating the properties of a class of nonequilibrium atom-surface interactions.
\end{abstract}

\date{\today}


\maketitle

\section{Introduction}

The existence of fluctuations of the electromagnetic field of both quantum and thermal origin gives rise to a force acting on neutral atoms in proximity to polarizable macroscopic bodies.
This force, which is the generalization of the van der Waals interaction, was theoretically investigated by Casimir and Polder \cite{Casimir48a}, and can be considered as the atomic counterpart of the Casimir force acting between macroscopic bodies  \cite{CasimirProcKonNederlAkadWet48}.
Because of its simple geometry, particular attention has been devoted to the interaction between an atom and a planar slab. For this configuration, the theoretical predictions have been confirmed by several experiments, based on different setups, including deflection of atomic beams close to surfaces \cite{deflection}, classical and quantum reflection of cold atoms \cite{Aspect1996,Shimizu2001,DeKieviet2003,Zao10} and Bose-Einstein condensates (BECs) \cite{Pasquini1,Pasquini2}, and dipole oscillations of BECs above dielectric substrates \cite{Cornell05,Cornell07}.

Deviations from the ideal case initially considered by Casimir and Polder, namely equilibrium systems at zero temperature and perfectly conducting plates, have been studied in the literature.
For instance, it was shown that the qualitative and quantitative features of Casimir-Polder forces can be tailored by acting on the optical properties of the surface \cite{Buhmann07,Rosa08,Pitaevskii08}.
In this context, an important role is played by surface resonances (such as surface plasmons in metals), i.e. collective coupled light-charge excitations propagating along the surface and exponentially decaying in the orthogonal direction.
These are known to give a relevant contribution in several fluctuation-induced phenomena, such as Casimir-related effects \cite{IntravaiaPRL} and radiative heat transfer \cite{Joulain}.
Moreover, the geometry of the slab can also play an important role, as it was recently shown by studying the interaction between an atom and a corrugated surface \cite{JPA,PRL,MessinaPRA,Reyes10,ImpensEPL10,Moreno1,Moreno2,Bimonte14,Bimonte15,Scheel15}.

Recently, it has been shown that placing the atom in an environment out of thermal equilibrium can modify the behavior of the force, giving rise to repulsive interactions and anomalous power-law dependencies \cite{Antezza,BuhmannPRL08,Cornell07,Buchleitner08,MessinaAntezza}.
Out-of-equilibrium configurations can be realized with different setups,
including configurations where different temperatures are simultaneously present in the system
(e.g. a different temperature of the slab and the environment~\cite{Antezza}) and also scenarios where one or more external lasers are acting on a system initially in thermal equilibrium.
In this last case the populations of one or more field's modes are modified, thus affecting the atom-field interaction.
In experiments lasers have already been used to actively tailor the potential felt by the atoms in order to produce atomic traps~\cite{Hammes03,Vetsch10} or to study the reflection of atomic beams \cite{Aspect1996,Shimizu2001,Bender10,Stehle11,BenderPRX14,Laliotis14,Laliotis15}, and
these methods were also employed to investigate the Casimir-Polder interaction in a variety of atom-surface distance regimes.

Our work is motivated by the opportunities offered by nonequilibrium configurations when coupled to material's optical properties and by the wide theoretical and experimental efforts in tailoring Casimir-Polder forces.
We start by studying how the atom-surface interaction energy is modified when the population of the surface-plasmon modes deviates with respect to its value at thermal equilibrium.
To this aim, we first theoretically assume a thermal imbalance between the surface-plasmon population and all the other modes, showing that this can modify the shape of the atomic potential giving rise, for instance, to a potential barrier.
Inspired by these results, we move to a scenario where one or two lasers undergo total internal reflection behind the surface, thus unbalancing the surface-plasmon population. By using a density-matrix description of the electromagnetic field, we show that the force produced by the laser(s) can be safely treated as an additive contribution with respect to the ordinary thermal-equilibrium Casimir-Polder force.

Our paper is structured as follows.
In Sec.~\ref{SecAtomSurface} we present the general formalism and the main ingredients of our calculation, stressing the role of the different contributions to the atom-surface interaction.
In Sec.~\ref{SecThermal} we discuss how a thermally unbalanced population of surface plasmons affects the Casimir-Polder interaction.
Sec.~\ref{SecLaser} is devoted to laser-modified interactions and provides analytic expressions of the one- and two-laser forces in terms of experimentally relevant parameters.
These expressions are used in Sec.~\ref{SecNumerics}, where we numerically explore how the atomic potential can be widely tailored in the presence of one or two external lasers. We show that both atomic barriers and traps can be produced.
We finally give in Sec.~\ref{SecConclusions} some final remarks.

\section{Atom-surface interaction}\label{SecAtomSurface}

In this section we briefly review some fundamental concepts of the physics of electromagnetic atom-surface fluctuation-induced interactions \cite{Casimir48a,Intravaia11}.
The interaction (free) energy of a neutral polarizable atom in an electromagnetic field is given by
\begin{equation}
U=-\frac{\langle \hat{\mathbf{d}}(t)\cdot \hat{\mathbf{E}}(\mathbf{R}_\text{a},t)\rangle}{2},
\label{GeneralInteractionEnergy}
\end{equation}
where the atom is modeled in terms of an electric dipole described by the operator $\hat{\mathbf{d}}$, and $\hat{\mathbf{E}}$ is the electric filed operator evaluated at the atom position $\mathbf{R}_\text{a}$.
The symbol $\langle\cdots\rangle$ denotes the mean value taken over the initial state of the atom + (field+matter) system of symmetrically ordered operators. Since $\hat{\mathbf{d}}$ and $\hat{\mathbf{E}}$ commute at equal times, this choice of ordering operators is irrelevant for the expression in Eq.~\eqref{GeneralInteractionEnergy}.
However, this is not always true when a perturbative approach is used.
In this case, a well-defined ordering choice has to be keep consistent throughout the calculation.
As it was pointed out by Dalibard, 
Dupont-Roc and Cohen-Tannoudji \cite{Dalibard82} the symmetric ordering 
is the only one that allows for a well defined physical interpretation
of the terms describing different system's contribution (see below).
Clearly, in equilibrium the final result will not depend on the specific choice and for nonequilibrium confgurations this allows
to avoid the appearance of unphysical terms.
We will assume that the state of the system is factorized and the corresponding density matrix is given as $\hat{\rho}= \hat{\rho}_{\rm a} \otimes\hat{\rho}_{\rm f}$, where both the field+matter (described by $\hat{\rho}_\text{f}$) and the atom (described by $\hat{\rho}_\text{a}$) are assumed to be locally in thermal equilibrium but not necessarily at the same temperature.

In a perturbative scheme, the dynamics of the dipole and field operators can be written as $\hat{\mathbf{d}}(t)\approx \hat{\mathbf{d}}^{\rm fr}(t)+ \hat{\mathbf{d}}^{\rm in}(t)$ and $\hat{\mathbf{E}}(\mathbf{R}_\text{a},t)\approx \hat{\mathbf{E}}^{\rm fr}(\mathbf{R}_\text{a},t)+ \hat{\mathbf{E}}^{\rm in}(\mathbf{R}_\text{a},t)$. The superscript ``fr'' indicates the free evolution, i.e., the dynamics of the atom without the field+matter system and vice versa. The ``in'' terms denote the contributions induced by the coupling, and in the frequency domain they can be written as
\begin{eqnarray}
\hat{\mathbf{d}}^{\rm in}(\omega) &=& \hat{\underline{\alpha}}(\omega)\cdot \hat{\mathbf{E}}^{\rm fr}(\mathbf{R}_\text{a},\omega), \nonumber \\
\hat{\mathbf{E}}^{\rm in}(\mathbf{R},\omega) &=& \underline{G}(\mathbf{R},\mathbf{R}_\text{a};\omega)\cdot \hat{\mathbf{d}}^{\rm fr}(\omega).
\label{InducedFreeRelations}
\end{eqnarray}
The quantity $\hat{\underline{\alpha}}(\omega)$ describes the atomic bare polarizability dyadic operator and $\underline{G}(\mathbf{R},\mathbf{R}_\text{a};\omega)$ is the electromagnetic Green tensor.
Within this perturbative approach, Eq.~\eqref{GeneralInteractionEnergy} can be then split into two contributions
\begin{equation}
U = -\frac{\langle \hat{\mathbf{d}}^{\rm in}(t)\cdot \hat{\mathbf{E}}^{\rm fr}(\mathbf{R}_\text{a},t)\rangle}{2} -
\frac{\langle \hat{\mathbf{d}}^{\rm fr}(t)\cdot \hat{\mathbf{E}}^{\rm in}(\mathbf{R}_\text{a},t)\rangle}{2} .
\label{SplittedInteractionEnergy}
\end{equation}
The first term is a ``field" term, that we call $U_\text{f}$.
It gives a contribution to the interaction energy arising from the intrinsic fluctuations of the field.
The second term, $U_\text{a}$, is an ``atom" term and contains the intrinsic fluctuations of the dipole.
Note that, within the symmetric order used here, both contributions are real and nonzero and the atom-surface interaction is therefore ascribable to both the atom's and the field's (quantum) fluctuations ~\cite{Milonni73,Dalibard82}
(an alternative ordering, normal or anti-normal, will attribute more weight to one of the previous terms).
If retardation is neglected and the field is treated classically, only the term $U_\text{a}$ is nonzero and it describes the van der Waals interaction between the atom and the surface \cite{London30}.
Therefore, it is commonly believed that for small enough atom-surface distances, $U_\text{a}$ provides the dominant attractive contribution to the interaction in the case of nonmagnetic systems.
Below we show, however, that this is not true in general, and that the strength as well as the sign of $U_\text{a}$ are strongly affected by the properties of the whole system.

\subsection{Atom and field contributions}

Considering that the atom and the field+matter subsystems are in two thermal states characterized by temperatures $T_\mathrm{a}$ and $T_\mathrm{f}$, respectively, the fluctuation-dissipation theorem yields \cite{Intravaia11}
\begin{equation}\label{GeneralU2}\begin{split}
U_\text{f}&=-\frac{\hbar}{2\pi}\int_0^{\infty}\hspace{-2mm}d\omega\,\coth\Bigl(\frac{\hbar\omega}{2k_{B}T_{\rm f}}\Bigr)
{\rm Tr}\bigl[\underline{\alpha}^{(T_{\rm a})}_{R}(\omega)\cdot\underline{\mathcal{G}}_{I}(\mathbf{R}_\text{a},\omega)\bigr],\\
U_\text{a}&=-\frac{\hbar}{2\pi}\int_0^{\infty}\hspace{-2mm}d\omega\,\coth\Bigl(\frac{\hbar\omega}{2k_{B}T_{\rm a}}\Bigr)
{\rm Tr}\bigl[\underline{\alpha}^{(T_{\rm a})}_{I}(\omega)\cdot\underline{\mathcal{G}}_{R}(\mathbf{R}_\text{a},\omega)\bigr].\\
\end{split}\end{equation}
Here, $\underline{\alpha}^{T_{\rm a}}$ is the atomic thermal polarizability tensor, and $\underline{\mathcal{G}}$ is the scattered part of the Green tensor,
$\underline{\mathcal{G}}(\mathbf{R}_\text{a},\omega)=\lim_{\mathbf{R}\to\mathbf{R}_\text{a}}\left[\underline{G}\left(\mathbf{R},\mathbf{R}_\text{a};\omega\right)-\underline{G}_0\left(\mathbf{R},\mathbf{R}_\text{a};\omega\right)\right]$,
where $\underline{G}_0$ is the free-space Green function. The subscripts ``$R$'' and ``$I$'' indicate the real and imaginary parts, respectively, while $\mathrm{Tr}$ indicates the tensorial trace operator. One can verify that by summing up the two contributions in Eq.~\eqref{GeneralU2},  the standard Casimir-Polder equilibrium ($T_\mathrm{a}=T_\mathrm{f}$) interaction energy is recovered.

The bare polarizability tensor operator appearing in Eq.~\eqref{InducedFreeRelations} is formally defined as
\begin{equation}
\hat{\underline{\alpha}}(\omega)=\frac{\rm i}{\hbar}\int_0^{\infty}dt[ \hat{{\bf d}}^{\rm fr}(t), \hat{{\bf d}}^{\rm fr}(0) ]  e^{{\rm i} \omega t}.
\end{equation}
In the case of a multi-level atom with states labeled by an integer index $n$ (each having energy $\hbar\omega_n)$, the elements of the tensor in the state $|n\rangle$ read
\begin{equation}
\alpha^{(n)}_{ij}(\omega)=\frac{2}{\hbar}\sum_{m\neq n}\frac{\omega_{mn}(\mathbf{d}_{mn})_i(\mathbf{d}_{mn})_j}{\omega_{mn}^2-(\omega+i0^ +)^2},
\end{equation}
where we add a small imaginary part in order to enforce causality and $\omega_{mn}=\omega_m-\omega_n$ is the transition frequency between states $m$ and $n$. For simplicity and without loss of generality, we also assumed that the matrix element $\mathbf{d}_{mn}=\langle m|\hat{{\bf d}}|n\rangle$ is real.  Starting from this expression, the thermal polarizability can be expressed as the weighted average
$ \alpha^{(T_\text{a})}_{ij}(\omega)=Z^{-1} \sum_n\alpha^{(n)}_{ij}(\omega)e^{-\frac{\hbar\omega_n}{k_\text{B}T_\text{a}}}$,
where $Z=\sum_n\exp(-\hbar\omega_n/k_\text{B}T_\text{a})$ is the partition function.
In this section we will focus on the case of a two-level isotropic atom with transition frequency $\omega_\text{a}$, for which we obtain the diagonal operator
$\alpha^{(T_\text{a})}_{ij}(\omega)= \delta_{ij} \alpha^{(T_\text{a})}(0) \omega_\text{a}^2/(\omega_\text{a}^2-\omega^2)$,
where $\alpha^{(T_\text{a})}(0)=\tanh(\hbar\omega_\text{a}/2k_\text{B}T_\text{a})\alpha_\text{g}(0)$ and $\alpha_\text{g}(0)$ is the ground-state static polarizability. Finally, using the identity
$\PP(1/x)=1/(x\pm{\rm i}\eta)\pm{\rm i}\pi\delta(x)$,
we have
\begin{equation}\label{AtomPol}
\underline{\alpha}^{(T_\text{a})}(\omega)=\alpha^{(T_\text{a})}(0)
\Bigl[\PP\Bigl(\frac{\omega_\text{a}^2}{\omega_\text{a}^2-\omega^2}\Bigr)+{\rm i}\frac{\pi\omega_\text{a}}{2}\delta(\omega-\omega_\text{a})\Bigr]\underline{1},
\end{equation}
where $\underline{1}$ is the identity operator.

Let us now move to the Green function. To this aim we consider the case of a semi-infinite homogeneous nonmagnetic medium occupying the region $z<0$, while the atom has positive $z$ coordinate. For this configuration the reflected part of the Green tensor appearing in Eq.~\eqref{GeneralU2} can be expressed as
\begin{equation}\label{Green}\begin{split}
\underline{\mathcal{G}}(\textbf{R},\omega)&=\frac{1}{8\pi\epsilon_0}\int_0^{\infty}dk\,k\kappa\Bigl[\Bigl(r_\text{TM}+\frac{\omega^2}{c^2\kappa^2}r_\text{TE}\Bigr)(\hat{\textbf{x}}\hat{\textbf{x}}+\hat{\textbf{y}}\hat{\textbf{y}})\\
&\,+2\frac{k^2}{\kappa^2}r_\text{TM}\hat{\textbf{z}}\hat{\textbf{z}}\Bigr]e^{-2\kappa z},
\end{split}\end{equation}
where we introduced the Fresnel reflection coefficients
$r_\text{TE}=(\kappa-\kappa_m)/(\kappa+\kappa_m)$ and $r_\text{TM}=(\epsilon\kappa-\kappa_m)/(\epsilon\kappa+\kappa_m)$.
Here, $\epsilon(\omega)$ is the dielectric permittivity of the medium,
$\kappa=\sqrt{k^2-\omega^2/c^2}$ and $\kappa_m=\sqrt{k^2-\epsilon \omega^2/c^2}$.
We see from Eq.~\eqref{Green} that, in virtue of the cylindrical symmetry of the problem, the interaction energy will only depend on the atom-surface distance $L=z_\text{a}$, i.e. on the $z$ coordinate of the atom. We will now focus on the specific case of a
metallic plate, whose dielectric properties can be modeled using the Drude model
$\epsilon(\omega)=1-\Omega_P^2/[\omega(\omega+{\rm i}\Gamma)]$,
where $\Omega_P$ is the plasma frequency and $\Gamma$ the relaxation rate.
This model predicts the existence of a surface plasmon mode, describing a polaritonic state of the field+matter system. Such a mode stems from collective oscillations of the electronic density of charge near the metal/vacuum interface and is associated to an evanescent field. The dispersion relation of the plasmon mode, existing only in TM polarization, can be derived from the poles of the transverse magnetic reflection amplitude $r_{\rm TM}$ at the interface. Hence, the general analytic expression of its dispersion relation can be derived by solving $\epsilon\kappa+\kappa_m=0$. In the simplified scenario of a lossless metal (i.e. $\Gamma=0$), the surface-plasmon dispersion relation follows straightforwardly and takes the form
\begin{equation}
\omega_{\rm sp}(k)=\Omega_{\rm P}\,\sqrt{\left(\!\frac{ck}{\Omega_{\rm P}}\!\right)^{\!\! 2}+\frac{1}{2} - \sqrt{\left(\!\frac{ck}{\Omega_{\rm P}}\!\right)^{\!\! 4}+\frac{1}{4}}}.
\label{sp_dispersion}
\end{equation}
For large values of $k$, the mode frequency tends to the asymptotic plasmonic frequency $\Omega_\text{sp}=\Omega_\text{P}/\sqrt{2}$.
When the atom is within the surface's near-field ($L\ll2\pi c/\Omega_\text{sp}$), the scattered Green tensor can be approximated by its electrostatic limit ($c\to \infty$).
If, in addition, we consider the lossless limit ($\Gamma\rightarrow0$) we can write,
\begin{equation}\begin{split}
\underline{\mathcal{G}}(L,\omega)&=\frac{\mathbf{\hat x \hat x}+ \mathbf{\hat y\hat y} + 2 \mathbf{\hat z\hat z}}{32\pi\epsilon_0L^{3}}\\
&\,\times\Bigl[\PP\Bigl(\frac{\Omega^{2}_{\text{sp}}}{\Omega^{2}_{\rm sp}-\omega^{2}}\Bigr)+{\rm i}\frac{\pi \Omega_{\text{sp}}}{2}\delta(\omega-\Omega_{\rm sp})\Bigr].
\end{split}\end{equation}

We are now ready to compute the two contributions to the interaction energy for an isotropic atom on top of a semi-infinite metallic medium, under the approximations stated above. We obtain
\begin{subequations}
\begin{equation}
U_\text{f}(L) = \frac{\hbar\Omega_{\rm sp}}{4}\frac{\coth{\left(\frac{\hbar\Omega_{\rm sp}}{2k_\text{B}T_\text{f}} \right)}}{8\pi\epsilon_0}
\PP\left(\frac{\omega_\text{a}^{2}}{\Omega_{\rm sp}^{2}-\omega_\text{a}^{2}}\right)\frac{\alpha^{(T_\text{a})}(0)}{L^3},
\label{energy-field-nonretarded}
\end{equation}
\begin{equation}
U_\text{a}(L) =-\frac{\hbar\omega_\text{a}}{4}
\frac{\coth{\left(\frac{\hbar\omega_\text{a}}{2k_\text{B}T_\text{a}}\right)}}{8\pi\epsilon_0}
\PP\left(\frac{\Omega^{2}_{\rm sp}}{\Omega^{2}_{\rm sp}-\omega_\text{a}^{2}}\right)
\frac{\alpha^{(T_\text{a})}(0)}{L^3} .
\label{energy-particle-nonretarded}
\end{equation}
\label{energies-nonretarded}
\end{subequations}
Note that, near the interface, the strength and the either attractive or repulsive nature of each of these two contributions depend on the relative magnitude of the atomic resonant frequency $\omega_\text{a}$ and the surface-plasmon frequency $\Omega_{\rm sp}$, as well as on the temperature of each subsystem:
The higher is $\Omega_{\rm sp}$ the less is the inertia of the charges in the medium.
As a result, when $\omega_\text{a}$ is much smaller than $\Omega_{\rm sp}$, the image dipole is perfectly anti-correlated with the atomic dipole, resulting in an attractive interaction. However, as soon as the atomic frequency approaches the surface plasmon frequency, the electrons in the metal cannot follow the high frequency oscillations of the dipole, leading to a reduction of the correlation with its image which results in a weaker force and eventually in a repulsive interaction. A dispersionless description for the material ignores a priori any internal dynamics of the medium.

To gain further insights it is interesting to contrast the previous result with that one would have obtained by using a constant positive permittivity $\epsilon=n^{2}>0$, where $n$ gives the material refractive index (this approach has been often used in the evaluation of the Casimir-Polder interaction). In this case one can show that in the near field this leads to $U_\text{f}(L)=0$ and to $U_\text{a}(L)$ given by the same expression
\eqref{energies-nonretarded}
but with the principal value term replaced by $r=(n^{2}-1)/(n^{2}+1)$.
As another example, let us consider the case of a surface made of an ideal metal, which can be
obtained by taking the limit $\Omega_P \rightarrow \infty$ in the previous expressions.
In this case the field term $U_\text{f}$ also vanishes, and the interaction is solely given by the atomic term $U_\text{a}$, which results monotonic and attractive.
It is well known that it can also be derived from an image dipole calculation, considering the interaction energy between the fluctuating dipole $\hat{\bf d}$ and the electric field $\hat{\bf E}_{\rm img}$ generated by its image $- \hat{\bf d}$ in the reflecting planar surface \cite{Aspect}.
Thus, one finds in this case
\begin{equation}
U_\text{a}(L)= - \frac{\langle \hat{\bf d} \cdot \hat{\bf E}_{\rm img} \rangle}{2}\propto-\frac{\hbar \omega_\text{a}}{4}\frac{\alpha_\text{g}(0)}{8 \pi\epsilon_0L^3}.
\end{equation}
However, for realistic materials showing a resonance in their permittivity, the field term is nonzero due to the existence of surface modes, which are absent otherwise.
In Eqs.~\eqref{energies-nonretarded} the field term can still be neglected with respect to the atomic term ($U_\text{f}/U_\text{a}\ll 1$) as long as $\omega_\text{a}\coth(\hbar\Omega_{\rm sp}/2k_\text{B}T_\text{f})\ll \Omega_{\rm sp}\coth(\hbar\omega_\text{a}/2k_\text{B}T_\text{a})$. For example, at thermal equilibrium $T_\text{f}=T_\text{a}=300$ K,
for a two-level atom with transition frequency $\omega_\text{a}=2.4\times10^{15}$\,rad/s (close to the two most relevant transitions of rubidium, see below)
in front of a gold surface (described using a Drude model with $\Omega_P=9\,$eV and $\Gamma=35\,$meV \cite{LambrechtEurJPhysD00}), $U_\text{f}/U_\text{a}=0.25$.

The two approximations performed here, namely a lossless metal and the quasi-static limit, give us a hint about the behavior of the field and atom contributions defined in Eq.~\eqref{GeneralU2} to the total atom-surface interaction.
In the end of this section, however, we abandon these approximations and study numerically the interaction at arbitrary atom-surface distances, at thermal equilibrium at zero temperature, between a two-level isotropic atom and a gold half-space. In order to perform this calculation, we first observe that $U_\text{f}$ and $U_\text{a}$ can be written as $U_\text{f}=\frac{U}{2}+\Delta$ and $U_\text{a}=\frac{U}{2}-\Delta$,
where $U$ is the total interaction energy at temperature $T$, and $\Delta$ is a correction that can be cast in the form
\begin{equation}\label{Delta}
\Delta=-\frac{\hbar}{4\pi}\int_0^{\infty}\hspace{-2mm}d\omega\,\coth{\Bigl(\frac{\hbar\omega}{2k_\text{B}T}\Bigr)}
{\rm Im}\,{\rm Tr}\bigl[\underline{\alpha}^{T*}(\omega)\cdot
\underline{\mathcal{G}}(\mathbf{R}_\text{a},\omega)\bigr] .
\end{equation}
It is well known (see e.g. \cite{Intravaia11}) that the total interaction $U$ at thermal equilibrium can be easily calculated by means of a Wick rotation, which at temperature $T=0$ gives
\begin{equation}
U(L)=-\frac{\hbar}{2\pi}\int_0^{+\infty}d\xi\,\alpha^{(0)}({\rm i}\xi){\rm Tr}\bigl[\underline{\mathcal{G}}(\mathbf{R}_\text{a},i\xi)\bigr].
\end{equation}
We tackle the correction $\Delta$ with the same approach. Hence, we start from the expression for the polarizability and we attribute to the pole at $\omega=\omega_\text{a}$ a small negative imaginary part, in order to be consistent with causality. Thus, the complex conjugate $\underline{\alpha}^{T*}(\omega)$ appearing in Eq.~\eqref{Delta} has a pole with positive imaginary part. This pole has to be taken into account when applying the residue theorem in order to move to the imaginary axis. Using this technique, it can be proven that
\begin{equation}
\Delta=\frac{U}{2}+\frac{\hbar\alpha_\text{g}(0)\omega_\text{a}}{4}{\rm Re}{\rm Tr}\bigl[\underline{\mathcal{G}}(\mathbf{R}_\text{a},\omega_\text{a})\bigr].
\end{equation}

\begin{figure}[t]
\includegraphics[width=0.47\textwidth]{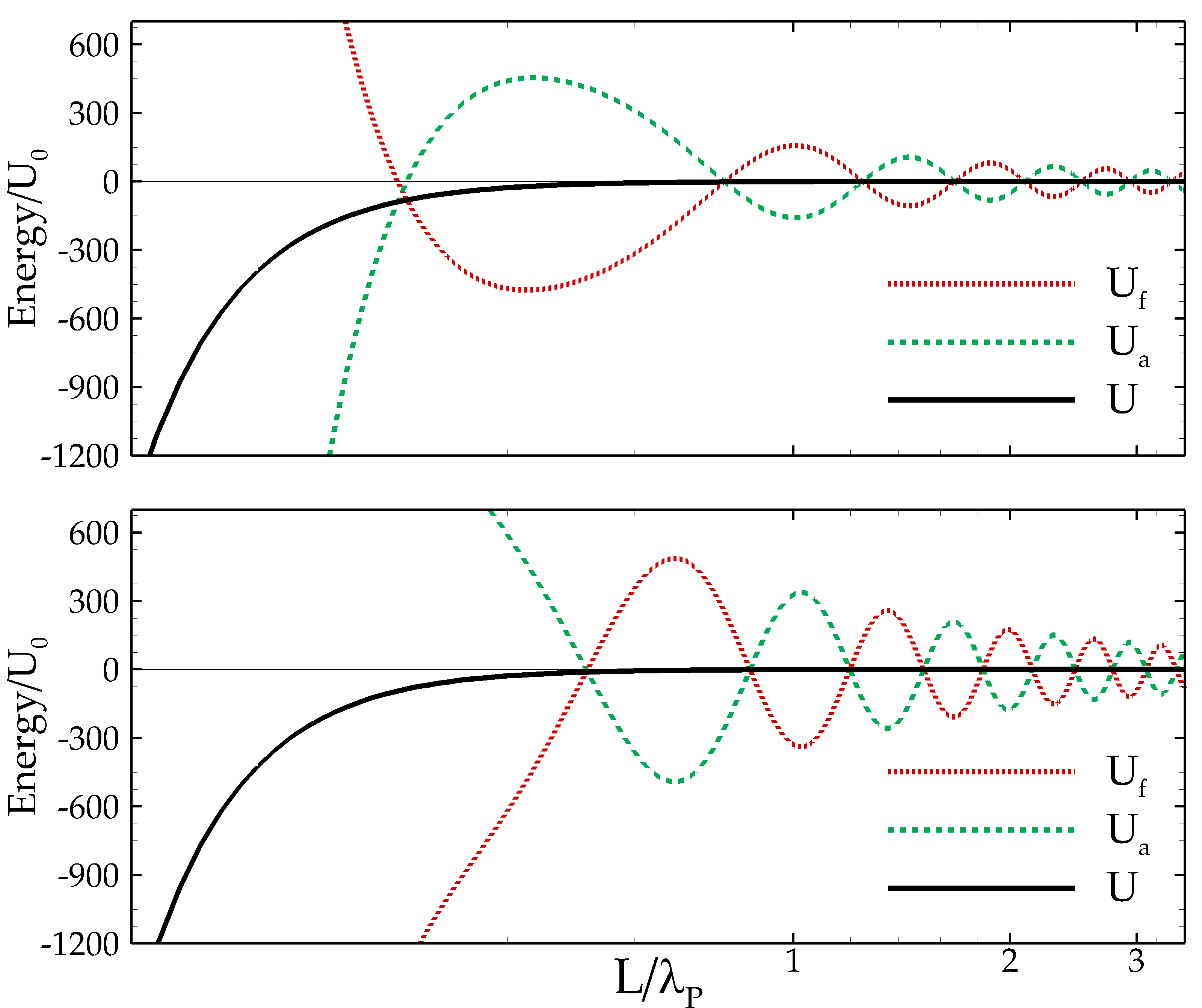}
\caption{(Color online) The different contributions to the interaction energy $U$ at $T_\mathrm{f}=T_\mathrm{a}=0$ for an isotropic atom above a gold half-space (Drude model with $\Omega_P=9\,$eV and $\Gamma=35\,$meV) as a function of the atom-wall distance $L$.
The (black) solid lines correspond to the total interaction energy $U$, the (red) dotted lines to $U_\text{f}$, and the (green) dashed lines to $U_\text{a}$. Distances are in units of the plasma wavelength  $\lambda_\mathrm{P}=2\pi c/\Omega_\mathrm{P}$, and energies are normalized to the total interaction energy $U_0$ at $L/\lambda_\mathrm{P}=1$. The top panel corresponds to a case where the atomic resonant frequency is below the
asymptotic plasmonic frequency $\Omega_{\rm sp}$ ($\omega_{\rm a}=0.85\,\Omega_\mathrm{sp}$), while the bottom panel corresponds
to the opposite situation ($\omega_{\rm a}=1.13\, \Omega_\mathrm{sp}$).}
\label{FigUfUa}
\end{figure}

In Fig.~\ref{FigUfUa} we plot the field and atom contributions (\ref{GeneralU2}) to the interaction energy as well as their sum at thermal equilibrium at $T=0\,$K as a function of distance, covering both the nonretarded and retarded regimes.  Although the total energy $U=U_\text{f}+U_\text{a}$ always corresponds to an attractive atom-surface force, each of the individual terms can correspond, at short distances, to attraction or repulsion, depending on the ratio $\Omega_{\rm sp}/\omega_{\rm a}$.
Both terms show oscillatory behaviors as a function of the atom-surface distance $L$ that are exactly out of phase. At large separations they cancel each other, while at short separations the attractive term dominates, resulting in the well known attractive and monotonic Casimir-Polder interaction.

\section{Thermal Imbalance Between Plasmonic and Other Modes}\label{SecThermal}

In the following, we concentrate on the contribution to the interaction energy arising from surface plasmon modes. In this section, in particular, we assess how the atom-surface interaction is modified in the presence of a thermal imbalance between the plasmons and the other modes of the electromagnetic field.

\begin{figure}[t]
\includegraphics[width=8cm]{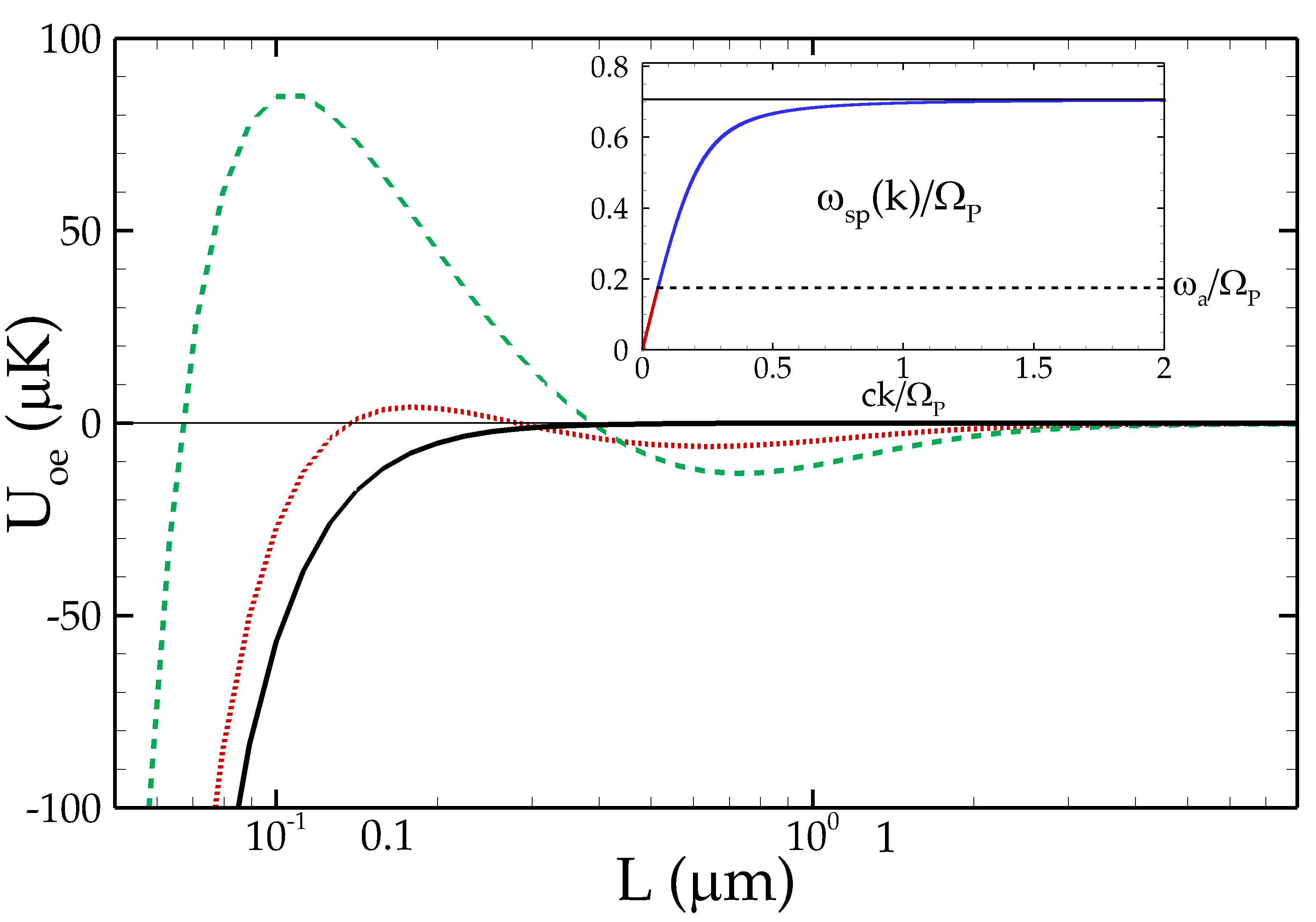}
\caption{(Color online) Out-of-equilibrium interaction energy $U_\mathrm{oe}$ for a rubidium atom above a gold half-space as a function of the distance for different thermal populations of the plasmonic branch.
The (black) solid line shows the thermal equilibrium situation ($T_\mathrm{sp} = T = 300\,\mathrm{K}$). Two imbalanced configurations with $T=300\,\mathrm{K}$ are also shown, corresponding to $T_\mathrm{sp} = 1100\,\mathrm{K}$ (red dotted) and $T_\mathrm{sp}=2000\,\mathrm{K}$ (green dashed). Inset: Dispersion relation $\omega_\mathrm{sp}(k)$ for the surface plasmon. Plasmonic modes with frequencies below (above) the atomic transition frequency $\omega_a$ give an attractive (repulsive) contribution to $U_\mathrm{oe}$.}
\label{FigThermicPlasmons}
\end{figure}

Let us, for simplicity, consider again a nondissipative ($\Gamma=0$) metallic plane, for which all mode resonances are real. The spectral decomposition of the trace of the Green dyadic is \cite{Markusevic88,Cole68}
\begin{equation}\label{spectral}\begin{split}
&\text{Tr}\bigl[\underline{\mathcal{G}}(L,\omega)\bigr]=\int_0^{+\infty}\!\!\!dk\sum_m {\cal R}_{k}(L,\omega_m)\\
&\,\times\Bigl[\PP\Bigl(\frac{2\omega_m}{\omega^2-\omega_m^2}\Bigr)+{\rm i}\pi\bigl(\delta(\omega+\omega_m)-\delta(\omega-\omega_m)\bigr)\Bigr],
\end{split}\end{equation}
where ${\bf k}$ ($k=|{\bf k}|$) is the component of the full wave-vector ${\bf K}$ parallel to the metal-vacuum interface, $\omega_m=\omega_m(k)$ is the dispersion relation for the $m$-th electromagnetic field mode (corresponding to a pole of the Green function, note that in this context we drop the $k$-dependence on the dispersion relation to enlighten the notation), and ${\cal R}_k(L, \omega_m)$ represents the corresponding residue. As discussed in Sec.~\ref{SecAtomSurface}, one of these modes corresponds to the surface-plasmon branch, whose dispersion relation is given in Eq.~\eqref{sp_dispersion} and is plotted in the inset of   Fig.~\ref{FigThermicPlasmons}. The dispersion relation \eqref{sp_dispersion} can be inverted  to give
\begin{equation}
k_{\rm sp}(\omega)=\frac{\omega}{c}\sqrt{\frac{\epsilon(\omega)}{\epsilon(\omega)+1}}=\frac{\omega}{c}\sqrt{\frac{\omega^2-\Omega_{\rm P}^2}{2\,\omega^2-\Omega_{\rm P}^2}},
\label{sp_dispersion_k}
\end{equation}
valid for $0<\omega<\Omega_{\rm sp}$. Equation \eqref{sp_dispersion_k} gives the value of $k$ required to excite a surface plasmon with a wave of frequency $\omega$ impinging on the metal/vacuum interface. In other words, it gives the angle of incidence needed to excite the surface plasmon at a given frequency.
The field and atom contributions to the interaction energy due to the plasmonic branch are
\begin{equation}\label{plasmon-field}\begin{split}
U_{\rm f,sp}(T_{\rm sp})&=\frac{\hbar\omega_a}{2}\alpha^{(T_\text{a})}(0)\,\,
\PP\int_0^{+\infty}\!\!\!dk\coth\Bigl(\frac{\hbar\omega_\text{sp}(k)}{2k_BT_\text{sp}}\Bigr)\\
&\,\times\frac{\omega_a}{\omega_a^2-\omega_{\rm sp}^2(k)} {\cal R}_{k}(L, \omega_{\rm sp}(k)),\\
U_{\rm a,sp}(T_{\rm a})&=-\frac{\hbar\omega_a}{2}\,\alpha^{(T_{\rm a})}(0)\coth\Bigl(\frac{\hbar\omega_a}{2k_BT_\text{a}}\Bigr)\\
&\,\times\PP\int_0^{+\infty}\!\!\!dk\frac{\omega_{\rm sp}(k)}{\omega_a^2-\omega_{\rm sp}^2(k)} {\cal R}_{k}(L, \omega_{\rm sp}(k)),
\end{split}
\end{equation}
where $T_a$ is, as usual, the atom temperature and $T_{\rm sp}$ is the temperature of the surface-plasmon mode, which can now be considered different from that of the other modes of the field.
To obtain an analytic expression for the residue ${\cal R}_{k}(L, \omega_{\rm sp}(k))$, we start by taking the trace of the integrand defining the Green function in Eq.~\eqref{Green}:
\begin{equation}\label{Fk}
\mathcal{F}_k(L,\omega)=\frac{k\kappa}{4\pi\epsilon_0}
\left[
\frac{\omega^2}{c^2\kappa^2}r_{\rm TE}+
\left(1+\frac{k^2}{\kappa^2}\right) r_{\rm TM}
\right] e^{-2\kappa L}.
\end{equation}
Hence, the residue is formally given by
\begin{equation}
{\cal R}_{k}(L, \omega_{\rm sp})=
\lim_{\omega\to\omega_{\rm sp}(k)}
\left[\omega-\omega_{\rm sp}(k)\right]\,\mathcal{F}_k(L,\omega).
\end{equation}
We remark that, since the plasmon frequency is a pole of the TM reflection amplitude, the term proportional to $r_\text{TE}$ can be neglected in Eq.~\eqref{Fk}.
As an alternative to the explicit calculation of the limit, one can recast $\mathcal{F}_k(L,\omega)$ as the ratio of a numerator and a denominator functions, denoted respectively by $\mathcal{N}_k(L,\omega)$ and $\mathcal{D}_k(L,\omega)$
which, in the specific case, result:
\begin{equation}\begin{split}
	&\mathcal{N}_k(L,\omega)=
	k \left(2 c^2 k^2-\omega ^2\right)
	e^{-\frac{2 L \sqrt{c^2 k^2-\omega ^2}}{c}}
	\\&\quad\times
	\left[
	\left(\omega ^2-\Omega_{\rm P}^2\right) \sqrt{c^2 k^2-\omega ^2}
	-\omega ^2 \sqrt{c^2 k^2-\omega ^2+\Omega_{\rm P}^2}\,
	\right],
	\\&\mathcal{D}_k(L,\omega)=
	4 \pi  c \epsilon_0 \sqrt{c^2 k^2-\omega ^2}
	\\&\quad\times
	\left[
	\left(\omega ^2-\Omega_{\rm P}^2\right) \sqrt{c^2 k^2-\omega ^2}
	+\omega ^2 \sqrt{c^2 k^2-\omega ^2+\Omega_{\rm P}^2}\,
	\right].
\end{split}\end{equation}
Note that solving in $\omega$ the condition $\mathcal{D}_k(L,\omega)=0$ gives the different mode branches, among which we find again the surface-plasmon dispersion $\omega_{\rm sp}(k)$ [Eq.~\eqref{sp_dispersion}].
Then, the residue can be more easily computed as
\begin{equation}
{\cal R}_{k}(L, \omega_{\rm sp}(k))=
\left.
\frac{\mathcal{N}_k(L,\omega)}
{\partial_\omega \mathcal{D}_k(L,\omega)}\right|_{\omega=\omega_{\rm sp}(k)},
\end{equation}
which, after lengthy but straightforward algebraic manipulations, gives
\begin{equation}\label{Residue}\begin{split}
{\cal R}_{k}(L,\omega_{\rm sp})
&=-\frac{\Omega_{\rm P}^3}{4\pi c^2\epsilon_0}
\frac{\omega_+(k) - \omega_-(k)}{\omega_{\rm sp}(k)}
\frac{\left(\frac{ck}{\Omega_{\rm p}}\right)^{\! 5} e^{-\frac{2L}{c}\omega_-(k)}}
{\sqrt{\left(\frac{ck}{\Omega_{\rm p}}\right)^4+\frac{1}{4}}},
\end{split}\end{equation}
with $\omega_\pm(k)=\Omega_{\rm P}\sqrt{\sqrt{\Bigl(\frac{ck}{\Omega_{\rm p}}\Bigr)^4+\frac{1}{4}}\pm\frac{1}{2}}$.
Equations~\eqref{plasmon-field} describe the full retarded plasmonic contribution to the atom/wall interaction energy and in the near field reduce to the expressions obtained in Eqs.~\eqref{energies-nonretarded}.

It is interesting to calculate how the total interaction energy is modified when the temperature in the plasmonic branch $T_{\rm sp}$ is different from that of the other electromagnetic modes and the atom \cite{Haakh2010}. Assuming that both the field and the atom are at temperature $T$, a change in $T_{\rm sp}$ only affects the field contribution in Eq.~\eqref{plasmon-field}.
Using the full equilibrium energy $U(T)$ at temperature $T$, the out-of-equilibrium interaction energy can be evaluated as
\begin{equation}
U_{\rm oe}(T,T_{\rm sp}) = U(T) - U_{\rm f,sp}(T) + U_{\rm f,sp}(T_{\rm sp}).
\label{ThermalPlasmonsTotalEnergy}
\end{equation}
In Fig.~\ref{FigThermicPlasmons} we show the behavior of $U_{\rm oe}$ as a function of the atom/wall distance for different values of $T_{\rm sp}$, assuming $T=300\,$K. For this calculation we take a two-level model for rubidium with transistion frequency $\omega_\text{a}=2.4\times10^{15}\,$rad/s and static polarizability $\alpha^{T_\text{a}}(0)/4\pi\epsilon_0\simeq46\times10^{-30}\,\text{m}^3$.
As $T_{\rm sp}$ is increased from the equilibrium value the energy becomes nonmonotonic, featuring a barrier at short separations and a stable minimum at larger distances. The origin of the crossover from attraction to repulsion can be explained as follows. Consider, at first, Eq.~\eqref{plasmon-field}. Since ${\cal R}_{k}(L,\omega_{\rm sp}(k))$ is real and negative [cf. Eq.~\eqref{Residue}], surface plasmons of frequency $\omega_{\rm sp}(k)<\omega_a$ give a negative (attractive) contribution, while for $\omega_{\rm sp}(k)>\omega_a$ this contribution is positive (repulsion, also see the inset of Fig.~\ref{FigThermicPlasmons}).
Considering the evanescent waves associated with the surface plasmons, it turns out that repulsive contributions ($\omega_{\rm sp}(k)>\omega_\text{a}$) correspond to fields with decay lengths shorter than those related to attractive contributions ($\omega_{\rm sp}(k)<\omega_a$). The plasmonic effects combine with the Casimir-Polder attraction, giving rise to the barrier at short distances while, farther from the wall, repulsive contributions become negligible and attraction takes over, resulting in a stable minimum (see Fig.~\ref{FigThermicPlasmons}).

Despite the previous interesting features, an experimental realization of the thermal imbalance described here might be, however, quite challenging. This would require an external source able to thermally excite a large number of plasmonic modes corresponding to different points along the dispersion relation given in Eq.~\eqref{sp_dispersion}. In Sec.~\ref{SecTwoLasers} we will investigate a more standard configuration where two lasers in the Kretschmann configuration \cite{Kretschmann68,Stehle11} are used.

\section{Laser-induced nonequilibrium effects}\label{SecLaser}

\begin{figure}[t]
\includegraphics[width=.9\linewidth]{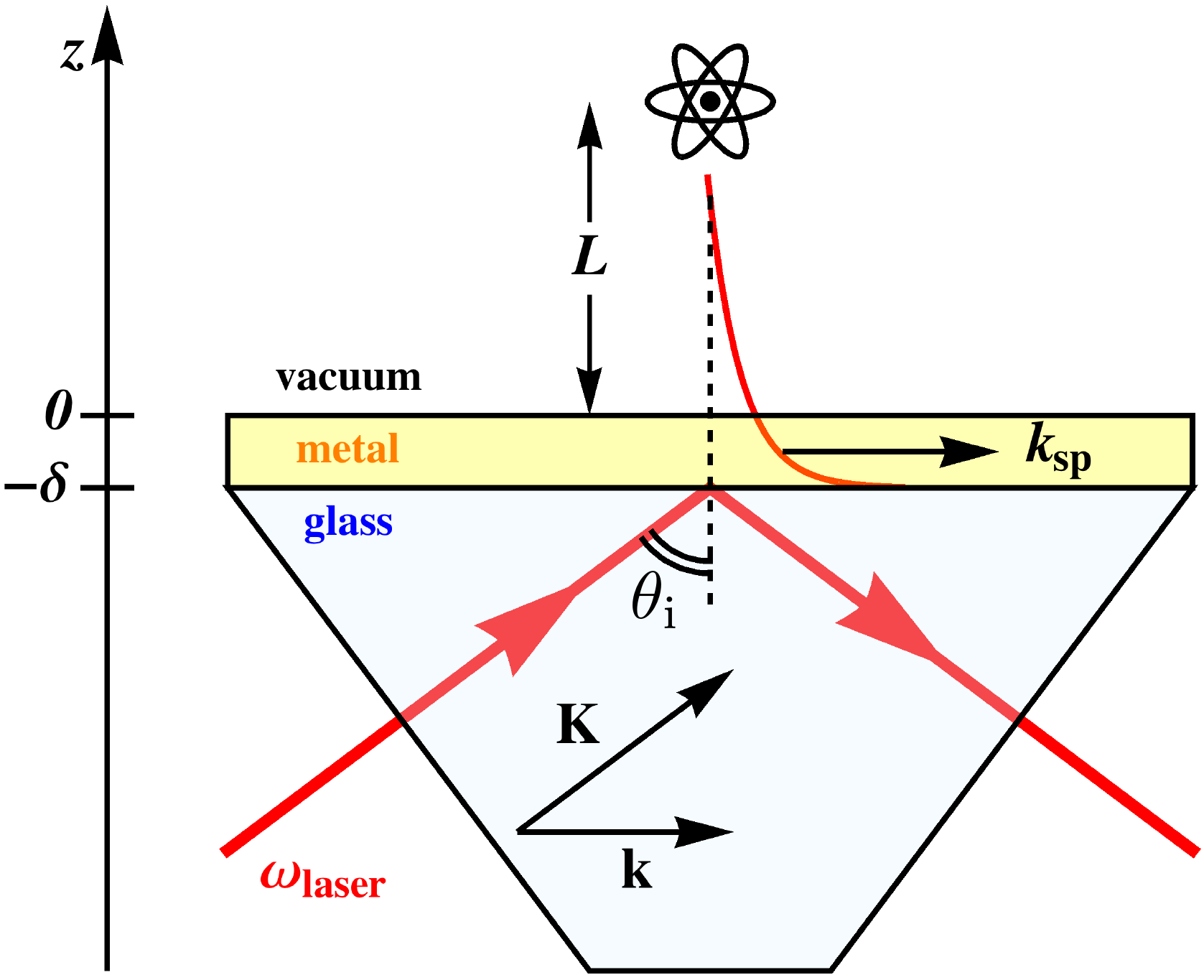}
\caption{(Color online) Kretschmann configuration of the system under investigation.}
\label{FigKretschmann}
\end{figure}

In Sec.~\ref{SecThermal} we showed that a change in the plasmonic population with respect to thermal equilibrium strongly affects the atom-surface interaction. A simple way to manipulate the population in the plasmonic branch is to use an external laser beam. One of the commonly used techniques involves the so-called Kretschmann configuration \cite{Kretschmann68} which is schematically illustrated in Fig.~\ref{FigKretschmann}. A laser beam with frequency $\omega_l$ propagates through a glass prism and undergoes total internal reflection on its upper face, which is coated with a thin metallic film.
Total internal reflection generates evanescent fields at the glass/metal interface and, when the projection of the laser wavevector on the plane, $k_l$, fulfills the condition $k_l=k_{\rm sp}(\omega_l)$ [cf. Eq.~\eqref{sp_dispersion_k}], surface plasmons are excited in the metal.
The combination of the ordinary Casimir-Polder interaction with the force due to the laser-induced evanescent electromagnetic field results in a tailorable nonequilibrium atom-surface interaction.
This well-known phenomenon has been used to measure the Casimir-Polder force itself \cite{Aspect1996} and, in more sophisticated configurations, to trap cold atoms close to a metallic surface \cite{Hammes03}.

In a recent experiment using the configuration of Fig.~\ref{FigKretschmann} the classical reflection of a rubidium Bose-Einstein condensate (BEC) from an energy barrier was measured \cite{Stehle11}. This barrier was formed by the superposition of the attractive Casimir-Polder force and a repulsive evanescent field generated by a laser blue-detuned with respect to the atomic transition.
The height of the energy barrier was estimated from the measured reflected BEC density versus the incoming kinetic energy. Varying the angle of incidence $\theta_{\rm i}$ of the laser on the metallic film, a maximum reflectivity was observed for an incident angle fulfilling the surface-plasmon excitation condition.
In \cite{Stehle11} a fitting model was used to explain the experimental results, where the Casimir-Polder interaction is simulated by the formula $U_{\rm CP}=-C_4/(L^3(L+\lambda_{0}))$, with the length scale $\lambda_{0}=780/2\pi$\,nm corresponding to the D2-line of rubidium. The plasmonic enhancement factor was numerically calculated from a multi-layer matrix model, taking into account the detailed composition of the glass/metal/vacuum interface.
The theoretical model used for this experiment is based on two approximations. The first is the description of the Casimir-Polder interaction as a power-law dependence modified by the inclusion of a length scale $\lambda_{0}$ in order to account for the transition from the near to the far zone, i.e. between the two regimes in which the atom-surface distance is small or large compared to a typical atomic-transition wavelength. The second approximation is the description of the interaction as the sum of the one produced by the laser and the standard Casimir-Polder potential.
The same additive approximation was used in a more recent work \cite{BenderPRX14}, where the CP interaction between an atom and a 1D grating was probed by means of diffraction of a Bose-Einstein condensate. In this case, the CP interaction was calculated exactly by using a Rayleigh decomposition describing the scattering upon a sapphire substrate on which gold stripes are deposited to form a grating. The modification to the evanescent repulsive potential due to the nanostructure was also studied using the same Rayleigh decomposition.

The theoretical results obtained in these two works are in good agreement with the experimental results. Nevertheless, additivity is indeed a delicate approximation, since the laser is affecting a mode of the electromagnetic field which is also providing a contribution to the CP interaction. In other words, the presence of the laser induces a nonequilibrium state of the electromagnetic field, for which the treatment of the CP interaction is known to be nontrivial \cite{Antezza,GorzaEurPhysJD06,Cornell07,BuhmannPRL08,SherkunovPRA09,Behunin,MessinaAntezza,Laliotis14,Laliotis15}.
One could argue that, since the laser field is produced by a source which is (statistically) independent from the vacuum and thermal fields present even in absence of the laser, the additive approximation is reasonably justified. We would like to provide in the next Section a more detailed discussion of this point, and to confirm  the additivity of the two interactions by means of a self-consistent approach.

\subsection{Additivity of laser-induced contributions}

In order to discuss the validity of the additivity approach discussed above, we start from a quantum treatment of the electromagnetic field, described in terms of a quantum operator and a density matrix.
Based on the discussion above, we first observe that the atomic contribution $U_{\rm a}$ is not modified by a change in the field state since it only depends on the atomic one.
In addition, since different field modes are to a good approximation independent, $U_{\rm f}$ is also unmodified except for the one affected by presence of the laser.
The field contribution of this mode to the CP interaction energy is given by
\begin{equation}\label{Eq.FieldTerm}
U_{\mu,\rm f}({\bf R}_\text{a})=-\frac{1}{2}\int_{-\infty}^\infty d\tau\,\alpha(\tau)\langle\hat{\bf E}_{\mu}({\bf R}_\text{a},t) \cdot \hat{\bf E}_{\mu}({\bf R}_\text{a},t-\tau)\rangle,
\end{equation}
where $\alpha(\tau)$ is the inverse-Fourier transform of the atomic polarizability (from now on we focus on the case of an isotropic atom). The average is taken over the state of the field mode, to which we can associate a density operator $\hat{\rho}_{\mu}$. The index $\mu$ denotes a set of quantum numbers that characterize the mode. The operator $\hat{\bf E}_{\mu}({\bf R}_\text{a},t)$ describes this single mode of the electric field with frequency $\omega_{\mu}$, and can be cast in the form
\begin{equation}
\hat{\bf E}_{\mu} (\mathbf{R},t) = {\rm i} \sqrt{2\pi\hbar\omega_{\mu}}
\left[ \hat{a}_{\mu}\,\mathbf{f}_{\mu}(\mathbf{R})\,e^{-{\rm i}\omega_{\mu} t} - \hat{a}_{\mu}^\dag\,\mathbf{f}_{\mu}^*(\mathbf{R})\,e^{{\rm i}\omega_{\mu} t}\right],
\end{equation}
where $\hat{a}_{\mu}\,\,(\hat{a}_{\mu}^\dag)$ is the usual annihilation (creation) operator of photons in the considered mode and $\mathbf{f}_{\mu}(\mathbf{R})$ is the mode-function, which includes the polarization vector (defined as in \cite{BartoloPRA12}) and depends on the geometry of the system.
For further simplification we write ${\bf f}_{\mu}({\bf R})=\hat{\bf e}_{\mu}\, f_{\mu}({\bf R})\, e^{{\rm i}\phi_{\mu}({\bf R})}$, where both $f_{\mu}({\bf R})$ and $\phi_{\mu}({\bf R})$ are real numbers. To simplify the notation, in the following we will omit the spatial dependency on the the atomic position ${\bf R}_\text{a}$. We also introduce the mode two-time correlation operator
\begin{equation}\label{Eq.Correlation}\begin{split}
\hat{\chi}_{\mu}(t,\tau)&=\hat{\bf E}_{\mu}(t)\cdot\hat{\bf E}_{\mu}(t-\tau)\\
&=2\pi\hbar\omega_{\mu} f^2_{\mu}\bigl[\hat{n}_{\mu}e^{{\rm i}\omega_{\mu}\tau}+(\hat{n}_{\mu}+1)e^{-{\rm i}\omega_{\mu}\tau}\\
&\,-\hat{a}_{\mu}^2 e^{{\rm i}\omega_{\mu}\tau} e^{2{\rm i}(\phi_{\mu}-\omega_{\mu} t)}-\hat{a}_{\mu}^{\dag 2} e^{-{\rm i}\omega_{\mu}\tau} e^{-2{\rm i}(\phi_{\mu}-\omega_{\mu} t)}\bigr],
\end{split}\end{equation}
where $\hat{n}_{\mu}=\hat{a}_{\mu}^\dag\hat{a}_{\mu}$ is the usual number operator. Expression \eqref{Eq.Correlation} can be used in Eq.~\eqref{Eq.FieldTerm} to obtain
\begin{equation}\label{Eq.FieldTerm2}\begin{split}
U_{\mu,\rm f}&=-\pi\hbar\omega\alpha(\omega_{\mu})f^2_{\mu}
\bigl[2\braket{\hat{n}_{\mu}+1/2}_\text{f}-\braket{\hat{a}_{\mu}^2}_\text{f}e^{2{\rm i}(\phi_{\mu}-\omega_{\mu} t)}\\
&\,-\braket{\hat{a}_{\mu}^{\dag 2}}_\text{f}e^{-2{\rm i}(\phi_{\mu}-\omega_{\mu} t)}\bigr],
\end{split}\end{equation}
where we used $\alpha(-\omega)=\alpha^*(\omega)=\alpha(\omega)$, a condition which is true if $\omega$ is far enough from an atomic resonance. The problem reduces, thus, to the evaluation of the mean value of the operators on the state $\hat{\rho}_{\mu}$, i.e. $\braket{\hat{O}}=\mathrm{Tr}(\hat{\rho}_{\mu}\hat{O})$.

A laser can be described in terms of a coherent state, i.e. as the eigenstate of the annihilation operator ($\hat{a}\ket{\alpha}=\alpha\ket{\alpha}$).
In general, the density operator describing any state of a defined field's mode can be written in terms of the so-called Glauber-Sudarshan $P$-representation as
\begin{equation}\label{Eq.DensityGeneral}
\hat{\rho}_{\mu}=\int d^2\alpha\,P(\alpha)\ket{\alpha}\bra{\alpha},
\end{equation}
where $d^2\alpha=d\,\text{Re}\alpha\,d\,\text{Im}\alpha$.
For a coherent state of complex amplitude $\beta$ it is given by $P_{\rm co}(\alpha)=\delta^{(2)}(\alpha-\beta)$,
while for a thermal state of temperature $T$ it is given by
$P_{\rm th}(\alpha)=\frac{1}{\pi\nu_{\mu}}{\rm exp}\left[-\frac{|\alpha|^2}{\nu_{\mu}}\right]$, where
$\nu_{\mu}=\braket{\hat{n}_{\mu}}_T=(e^{\hbar\omega_{\mu}/k_\text{B}T}-1)^{-1}$, i.e. the bosonic thermal occupation number for that specific mode.

The key point we need at this stage is the expression of the $P$-representation of a field resulting from the superposition of two contributions.
More specifically, we are looking for the $P$-representation of a one-mode field associated to the simultaneous presence of an external laser and of a thermal contribution.
According to the description give by Lachs \cite{Lachs65}, the $P$ representation for the state of a field's mode in the presence of two arbitrary contributions is given by the convolution
\begin{equation}
P(\alpha)=\int d^2\beta\,P_1(\beta)P_2(\alpha-\beta) ,
\end{equation}
where $P_1(\alpha)$ and $P_2(\alpha)$ are the $P$-representations of these two contributions, respectively.
We assume  that $P_2$ corresponds to a thermal field and $P_1$ is left unspecified for the moment.
As seen before, to obtain the CP interaction we need the mean values $\langle\hat{n}_{\mu}+1/2\rangle_P$ and $\langle\hat{a}_{\mu}^2\rangle_P$ on the global single-mode state described by $P$.
For the former we have
\begin{equation}\label{EqAdd}\begin{split}
\langle\hat{n}_{\mu}+1/2\rangle_P&=\iint d^2\alpha\,d^2\beta\,P_1(\beta)P_{\rm th}(\alpha-\beta)\Bigl(|\alpha|^2+\frac{1}{2}\Bigr)\\
&=\int d^2\beta\,P_1(\beta)\Bigl(|\beta|^2+\nu_{\mu}+\frac{1}{2}\Bigr)\\
&=\langle\hat{n}_{\mu}\rangle_{P_1}+\nu_{\mu}+\frac{1}{2},
\end{split}\end{equation}
where $\langle\cdots\rangle_{P_1}$ describes a mean value taken on the state described by $P_1$.
For $\hat{a}_{\mu}^2$ and its hermitian conjugate (whose average values on the thermal state vanish) we just have
$\langle\hat{a}_{\mu}^2\rangle_P=\langle\hat{a}_{\mu}^2\rangle_{P_1}$.
This shows that the thermal component of the state will always be completely decoupled from the effects of the superimposed state of the field, whatever this state is.
This is indeed a result connected to the specific peculiar properties of the thermal state (mathematically speaking, to its Gaussian shape). It shows that in the CP interaction quantum ``interference effects'' cannot occur with the thermal component.
We remark from Eq.~\eqref{EqAdd} that the vacuum contribution (corresponding to the term $1/2$) has to be taken into account only once. In other words, if we correctly interpret the entire term $\nu_{\mu}+1/2$ as the one giving the standard CP interaction in thermal equilibrium, we only need to add the average value of $\hat{n}_{\mu}$ and not of $\hat{n}_{\mu}+1/2$ on the state $P_1$ of the superimposed field.
Finally, the contribution to the interaction associated with this specific mode is the sum of the ordinary thermal one and of $\widetilde{U}_{\mu,P_1}$, given by
\begin{equation}\begin{split}
\widetilde{U}_{\mu,P_1}&=-\pi\hbar\omega\alpha(\omega_{\mu})f^2
\bigl[2\braket{\hat{n}_{\mu}}_{P_1}-\braket{\hat{a}_{\mu}^2}_{P_1} e^{2{\rm i}(\phi_{\mu}-\omega_{\mu} t)}\\
&\,-\braket{\hat{a}_{\mu}^{\dag 2}}_{P_1} e^{-2{\rm i}(\phi_{\mu}-\omega_{\mu} t)}\bigr].
\end{split}\end{equation}
Thus, $\widetilde{U}_{\mu,P_1}$ is the contribution we need to add to the total equilibrium atom-surface interaction $U(T)$ in order to get the correct out-of-equilibrium result $U_{\rm oe}=U(T)+\widetilde{U}_{\mu,P_1}$.

The previous discussion provides a self-consistent justification of the additive treatment of the laser interaction and the CP thermal potential, as previously used in \cite{Stehle11,BenderPRX14}.

\subsection{One- and two-laser force}

In this subsection we specify the treatment introduced above to the case in which one or two external lasers perturb one or two modes of the field.
In the one-laser case, the state $P_1$ has to be replaced with the coherent state $P_{\rm co}(\beta)$.
We identify its amplitude by the complex number $\beta=|\beta|e^{{\rm i}\zeta}$.
The mean values that we need are
$\braket{\hat{n}_{}}_{\rm co}=|\beta|^2$ and
$\braket{\hat{a}^2}_{\rm co}=\braket{\hat{a}^{2\dag}}_{\rm co}^*=|\beta|^2 e^{2{\rm i}\zeta}$.
It follows that the coherent contribution to the interaction that has to be added to the standard CP potential reads (from now on we omit the explicit dependence to the mode $\mu$)
\begin{equation}\label{Eq.EnergyC}
\widetilde{U}_{\rm co}=-2\pi\hbar\omega\alpha(\omega_l) f^2
|\beta|^2\Bigl[1-\cos[2(\phi+\zeta-\omega_l t)]\Bigr].
\end{equation}
We observe that this contribution goes to zero in absence of the laser, i.e. for $\beta=0$.

In order to evaluate numerically Eq.~\eqref{Eq.EnergyC} for a specific experimental configuration,
we need to relate the amplitude $\beta$ to the parameters of the laser impinging on the structure represented in Fig.~\ref{FigKretschmann}.
To this aim, we start by considering the modes of the electromagnetic field at the interface between a dielectric semi-infinite medium $z<0$ described by a real refractive index $n$, and the vacuum $z\ge0$).
This case has been studied by Carniglia and Mandel \cite{Carniglia71,Birula72}, who
quantized the field in the two half-space geometry. The modes of a dielectric-vacuum planar geometry are triplets of impinging, reflected, and transmitted plane waves.
For a wave impinging the interface from the dielectric to the vacuum, the triplet is \cite{BartoloPRA12}
\begin{equation}\label{fmodes}
{\bf f}(\omega,\theta_{\rm i},{\bf r})=\mathcal{N}\,\hat{\textbf{e}}
\begin{cases}
e^{{\rm i} {\bf k \cdot r}}\,e^{{\rm i} k_{zd}z}+r_{\rm int}\,  e^{{\rm i} {\bf k \cdot r}}\,e^{-{\rm i} k_{zd}z}
& z<0,\\
t_{\rm int}\,  e^{{\rm i} {\bf k \cdot r}}\,e^{{\rm i} k_{z}z} & z\ge0,
\end{cases}
\end{equation}
where $\mathbf{r}$ is the projection of $\mathbf{R}$ on the plane of the interface. In the previous expression, $k_z$ ($k_{zd}$) is the $z$ component of the wavevector in vacuum (in the dielectric),  $r_{\rm int}$ and $t_{\rm int}$ are the reflection and transmission coefficients of the dielectric-vacuum interface, $\hat{\textbf{e}}$ is the polarization vector, and $\mathcal{N}$ is a dimensional normalization constant.
The wavevector components are linked to the ones defined in Sec.~II by $k_z={\rm i}\kappa$ and $k_{zd}={\rm i}\kappa_d$.
The Kretschmann configuration of Fig.~\ref{FigKretschmann} differs from the Carniglia-Mandel dielectric-vacuum configuration by the presence of a metallic spacer layer of thickness $\delta$.
As a consequence, the transmission and reflection coefficients to be used are the ones of a glass-metal-vacuum structure, which of course depend on the thickness $\delta$. Since in the following we only need the field in the vacuum region, we will make explicit use of the transmission coefficient of the structure
\begin{equation}\label{tstr}
t_\text{str}=\frac{t_\text{gl/Au}t_\text{Au/vac}e^{-\kappa_\text{Au}\delta}}{1+r_\text{gl/Au}r_\text{Au/vac}e^{-2\kappa_\text{Au}\delta}},
\end{equation}
where $r_{i/j}$ ($t_{i/j}$) represents the ordinary Fresnel reflection (transmission) coefficient from medium $i$ to medium $j$.
In the following we consider only TM polarization because this is the one responsible for the excitation of surface plasmons in the metallic layer. The vector $\bf k$ is the wave vector on the interface plane $z=0$, spanned by $\bf r$. The wave vector in the $z$ direction changes when passing from the medium to the vacuum. Details can be found in \cite{BartoloPRA12}, but the relevant point is that everything can be written in terms of the wave frequency $\omega$ and incidence angle $\theta_{\rm i}$ (see Fig.~\ref{FigKretschmann}).

We now determine the normalization factor $\mathcal{N}$. For the case of the laser mode described by a coherent state of amplitude $\beta$, we have that
$\langle\beta| {\hat E}^2|\beta\rangle= 2 \pi \hbar \omega (2 |\beta|^2 +1) |{\bf f}|^2
\simeq 4 \pi \hbar \omega |\beta|^2 |{\bf f}|^2$,
where we neglected fast oscillating terms. Therefore, within the glass right before the interface with the metal layer, the impinging electric field is such that
$E_{\text{gl}}^2=4 \pi \hbar \omega |\beta|^2 \mathcal{N}^2$,
and  the intensity of the laser beam in the glass region is related to this value by
\begin{equation}
\frac{c\epsilon_0}{2}n_{\text{gl}}\, E_{\text{gl}}^2=I_{\text{gl}}=\frac{P_l}{2\pi w_l^2},
\end{equation}
where $n_{\rm gl}(\omega_l)=\sqrt{\varepsilon_{\rm gl}(\omega_l)}$ is
the refractive index of the glass and $w_l$ is the waist of the laser beam within the glass. The laser-pumped mode beyond the metallic layer is, hence, entirely defined in terms of experimental quantities. The expression of these quantities as a function of the laser parameters in vacuum is straightforward (see e.g. \cite{Bender10}). In conclusion, the expression of the one-laser contribution to the atomic potential reads
\begin{equation}
\label{EqOneLaserPotential}
\begin{split}
\widetilde{U}_{\text{co}}(\theta_{\rm i},z)&=-\frac{2P_l}{c\,n_{\rm gl}(\omega_l)\,w_l^2}\,\frac{\alpha(\omega_l)}{4\pi\varepsilon_0}\,
\left|t_{\rm str}[\omega_l,k(\theta_{\rm i})]\right|^2\\
&\,\times e^{-2\,\kappa[\omega_l,k(\theta_{\rm i})]L}.
\end{split}\end{equation}
The dependence on $\theta_{\rm i}$ appears through $k$, i.e. the component of the laser-mode wave vector parallel to the interface. Since the laser beam is traveling in the glass, the modulus of the total wave vector is $K=n_{\text{gl}}(\omega_l)\omega_l/c$.
It follows that $k(\theta_{\rm i})=n_{\rm gl}(\omega_l)\omega_l\sin(\theta_{\rm i})/c$.
The latter gives explicitly the $\theta_{\rm i}$ dependence of $k$ needed in Eq.~\eqref{EqOneLaserPotential}.

Let us now move to the case of the interaction energy in the presence of two lasers, labeled with an index $j=1,2$. The two lasers have frequencies $\omega_j$, waists $w_j$, and incidence angles $\theta_j$. They are described in terms of coherent states having amplitudes $\beta_j=|\beta_j|e^{{\rm i}\zeta_j}$. Moreover, we will denote with $t_j$ the transmission coefficient \eqref{tstr} calculated at the wavevector and frequency of each laser. In order to deduce the energy we start with the two-mode (2M) electric field, which reads
\begin{equation}
\hat{\bf E}_{\rm 2M} (\mathbf{R},t) =\sum_{j=1,2}
{\rm i} \sqrt{2\pi\hbar\omega_j}
\Bigl[ \hat{a}_j\,\mathbf{f}_j(\mathbf{R})\,e^{-{\rm i}\omega_j t} - \hat{a}^\dag_j\,\mathbf{f}_j^*(\mathbf{R})\,e^{{\rm i}\omega_j t}\Bigr].\end{equation}
After similar calculations as the ones done for the one-laser case, and neglecting fast-oscillating terms, one gets the two-laser coherent contribution
\begin{widetext}
\begin{equation}
\label{EqTwoLaserGeneral2}
\begin{split}
\widetilde{U}_{\text{2M}}&=
-\sum_{j=1,2}\frac{2\,P_j}{cn_{\rm gl}(\omega_j)w_j^2}\,\frac{\alpha(\omega_j)}{4\pi\varepsilon_0}|t_j|^2e^{-2\kappa_jL}\\
&\,-\frac{2\sqrt{P_1\,P_2}}{c\sqrt{n_{\rm gl}(\omega_1)n_{\rm gl}(\omega_2)}w_1w_2}\frac{\alpha_1+\alpha_2}{4\pi\varepsilon_0}|t_1t_2|\,e^{-(\kappa_1+\kappa_2)L}\,(\hat{\textbf{e}}_1\cdot\hat{\textbf{e}}_2)\,\cos\bigl[\Delta\phi+\Delta\zeta-(\omega_1-\omega_2)\,t\bigr],
\end{split}
\end{equation}
\end{widetext}
where $\Delta\phi=\phi_1-\phi_2$ and $\Delta\zeta=\zeta_1-\zeta_2$.
We remark that the factor $\phi$ giving the phase of the mode function $\textbf{f}_j$ comes, in general, both from the transmission coefficient $t_j$ and the position-dependent term $e^{{\rm i}\mathbf{k}\cdot\mathbf{r}}$.

Finally, we consider the specific case of two lasers having the same power ($P_1=P_2=P_l$), frequency ($\omega_1=\omega_2=\omega_l$), waist ($w_1=w_2=w_l$), incidence angle
($\theta_1=\theta_2=\theta_{\rm i}$), but opposite wavevectors satisfying
\begin{equation}
\mathbf{k}_1=-\mathbf{k}_2=\frac{n_\text{gl}(\omega_l)\omega_l}{c}\sin(\theta_{\rm i})\hat{\mathbf{x}}.
\end{equation}
It follows that the two transmission coefficients $t_1=t_2=t_l$ coincide and thus $\phi_1=-\phi_2=\omega_ln_\text{gl}(\omega_l)\sin(\theta_{\rm i})/c$. Taking for the TM polarization unit vectors the usual definition $\hat{\mathbf{e}}=c/\omega(-k\hat{\mathbf{z}}+k_z\hat{\mathbf{k}})$, we conclude that
\begin{eqnarray}\label{EqTwoLaserCounter}
\widetilde{U}_{\text{2M}}&=&-\frac{4P_l}{cn_{\rm gl}(\omega_l)w_l^2}\frac{\alpha(\omega_l)}{4\pi\varepsilon_0}|t_l|^2\,e^{-2\kappa L} \nonumber \\
&& \times
\Bigl[1-\cos(2\theta_{\rm i})\cos\Bigl(2\sin(\theta_{\rm i})\frac{n_{\rm gl}(\omega_l)\omega_l}{c}x\Bigr)\Bigr] .
\end{eqnarray}
For sake of simplicity we also assumed $\Delta\zeta=0$, but we remark that a phase difference between the coherent states $\Delta\zeta\neq0$ simply leads to a shift of the potential \eqref{EqTwoLaserCounter} on the $x$ axis. From this expression we see that the presence of two counter-propagating lasers having the same frequency naturally introduces a spatial modulation of the interaction energy. We will numerically investigate this dependence in Sec.~\ref{Twolasers}.

\section{Numerical investigations}\label{SecNumerics}

This Section presents some numerical result showing how the aforementioned out-of-equilibrium effects would appear in experimentally realizable systems. In what follows, we refer to the Kretschmann setup sketched in Fig.~\ref{FigKretschmann}. We consider a sapphire prism and model its optical properties on the base of experimental data \cite{Palik98}. The metallic layer is a $50$~nm thick film of gold, whose dielectric properties are described by the Drude model.
We consider a $^{87}$Rb atom, which presents two main ground-to-excited state transitions having frequencies and matrix elements of the dipole operator $\omega_1=23.6943\times10^{14}$\,rad/s, $\omega_2=24.1419\times10^{14}$\,rad/s, $d_1=25.377\times10^{-30}$\,C~m and $d_2=35.842\times10^{-30}$\,C~m. To make our model more realistic, we also associate a finite line width to the resonances by using the values $\gamma_1=36.1283\times10^6$\,rad/s and $\gamma_2=38.1201\times10^6$\,rad/s \cite{SteckData}. Hence, we implement the isotropic polarizability of rubidium as
\begin{equation}\label{EqRbPolarizability}
\alpha_{\rm Rb}(\omega)=\frac{1}{4\pi\epsilon_0}\frac{2}{3\hbar}\sum_{i=1,2}\frac{\omega_i\,d_i^2}{\omega_i^2-(\omega+{\rm i}\gamma_i)^2},
\end{equation}
thus neglecting the thermal population of the excited states, which is an excellent approximation at room temperature for electric transitions. We stress here that Eq.~\eqref{EqRbPolarizability} will be used to calculate the Casimir-Polder interaction, whereas we will only consider its real part when evaluating the laser contributions to the interaction. This is justified by the fact that we will consider laser frequencies far enough from atomic resonances such that the imaginary part of $\alpha_\text{Rb}$ is negligible.

\begin{figure}[t]
\includegraphics[width=0.47\textwidth]{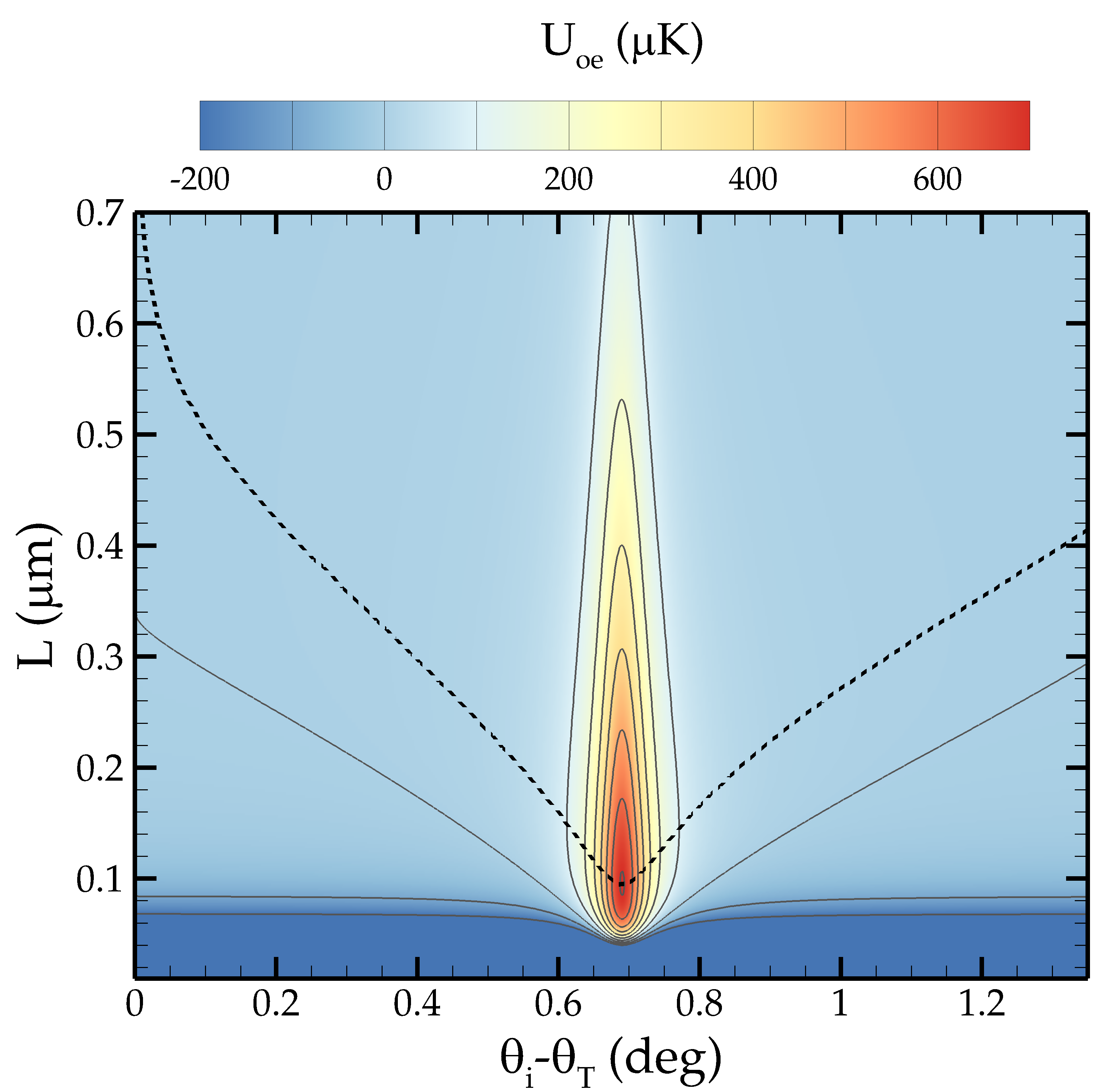}
\caption{\label{FigFlame} Total out-of-equilibrium interaction energy $U_{\rm oe}$ for a rubidium atom in front of a plasmon-excited glass/gold structure (configuration of Fig.~\ref{FigKretschmann}) at $T=300$~K. The contour plot shows the dependence on the atom-surface distance $L$ and on the incidence angle of the excitation laser $\theta_{\rm i}$. From the latter we subtract the total internal reflection angle $\theta_{\rm T}$ for the bare glass/vacuum interface. The frequency of the laser is $\omega_l=24.6\times10^{14}$~rad/s, which implies $\theta_{\rm T}\simeq34.23$~deg. The laser is blue-shifted with respect to the main transitions of rubidium, whose polarizability is modeled according to Eq.~\eqref{EqRbPolarizability}. The power and waist of the beam within the glass are, respectively, $P_l=200$~mW and $w_l=180$~$\mu$m.	The thick dashed curve marks the position of the potential-barrier maximum height as a function of $\theta_{\rm i}$.
The other curves are contour lines.}
\end{figure}

\subsection{One-laser force}

Let us begin by a one-laser configuration like the one of Fig.~\ref{FigKretschmann}. In order to have a repulsive evanescent contribution, we choose a laser frequency $\omega_l=24.6\times10^{14}$~rad/s, i.e. blue-shifted with respect to both rubidium's main transitions. In Fig.~\ref{FigFlame} we show the dependence of the total interaction energy on the atom-surface distance $L$ from the metal/vacuum interface and on the angle of incidence $\theta_{\rm i}$ of the laser beam at the glass/metal interface. While the Casimir-Polder interaction is always attractive in the absence of the laser, here a potential barrier appears. The maximum height of this barrier is greatly increased when $\theta_{\rm i}$ is such that Eq.~\eqref{sp_dispersion_k} is satisfied, corresponding to the excitation of surface plasmons which amplify the evanescent field outside the metallic layer. For realistic power and beam waist of the laser, we estimate a maximum barrier height of about $700$~$\mu$K.

\begin{figure}[t]
\includegraphics[width=0.47\textwidth]{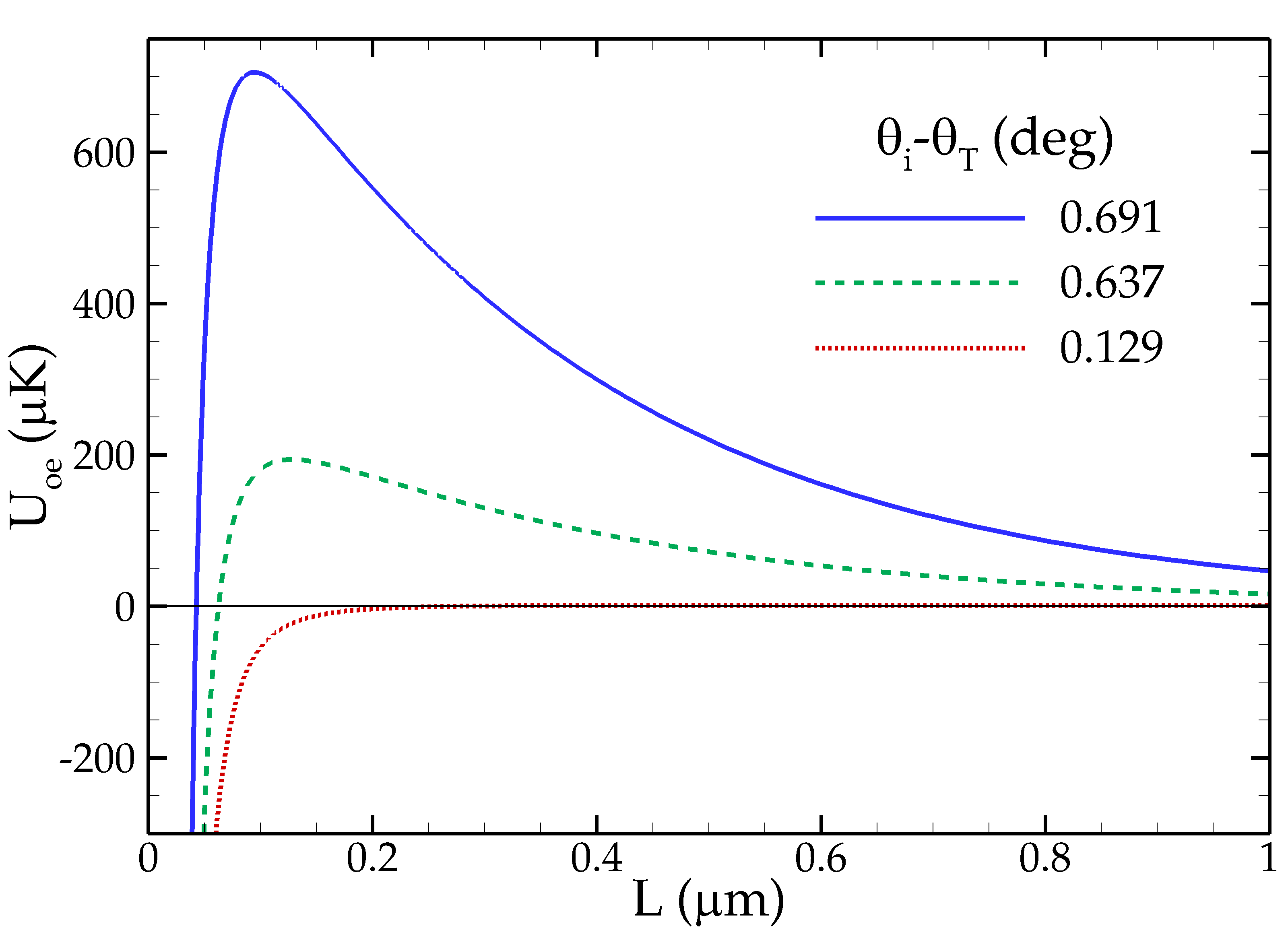}
\caption{\label{FigBarriers} Total out-of-equilibrium atom-surface interaction energy $U_{\rm oe}$ as a function of $L$ for the same configuration of Fig.~\ref{FigFlame}. Results for three different values of the incidence angle $\theta_{\rm i}$ of the exciting laser beam are shown, corresponding to vertical slices in Fig.~\ref{FigFlame}.}
\end{figure}

The behavior of the total potential as a function of the atom/interface distance for a fixed angle of incidence is shown in Fig.~\ref{FigBarriers}. The angles considered are all beyond the total internal reflection one $\theta_{\rm T}$, so that an evanescent-wave contribution is always present. We can appreciate how, far from the plasmonic resonance, such a contribution is fairly negligible, leaving the total potential mainly attractive. Instead, a barrier rapidly grows in the vicinity of the plasmonic resonance. Anyhow, the Casimir-Polder attraction eventually takes over the evanescent repulsion for distances below 100~nm.

\begin{figure}[t]
\includegraphics[width=0.47\textwidth]{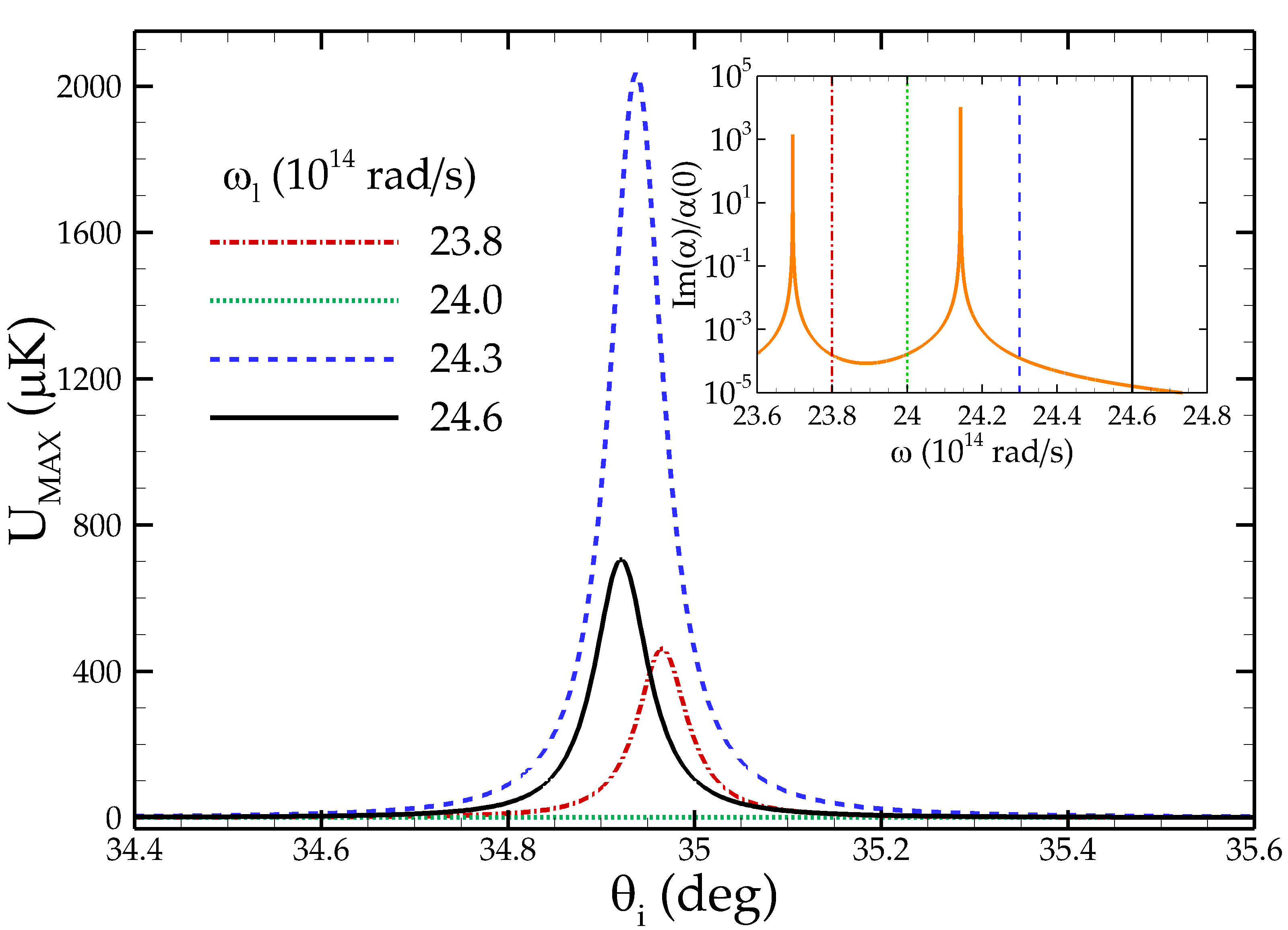}
\caption{\label{FigOmega} Maximum height of the atom-surface potential barrier $U_{\rm MAX}$ as a function of the exciting-laser incidence angle for different values of the laser frequency (see legend). Inset: Imaginary part of the polarizability $\alpha_{\rm Rb}$ [Eq.~\eqref{EqRbPolarizability}] over its zero-frequency value $\alpha(0)$ as a function of the frequency $\omega$. The four vertical lines mark the laser frequencies used in the main plot.}
\end{figure}

In Fig.~\ref{FigOmega} we investigate the behavior of the barrier maximum as a function of $\theta_{\rm i}$ for different frequencies $\omega_l$ of the impinging laser.
The latter turns out to be a critical parameter in the determination of the maximum barrier height. Each $\omega_l$ can be blue- or red-shifted with respect to each rubidium line (cf. inset of Fig.~\ref{FigOmega}). If the strongest coupling is with the transition above $\omega_l$, the overall evanescent contribution is attractive and no barrier arises (e.g. $\omega_l=24.0\times10^{14}$~rad/s in the figure). Otherwise, the barrier height can be hugely enhanced by taking $\omega_l$ slightly above a rubidium transition.
This effect is clearly due to the resonances in $\alpha_{\rm Rb}$ [Eq.~\eqref{EqRbPolarizability}], which appears as a prefactor of the evanescent contribution \eqref{EqOneLaserPotential}. The amplification cannot be pushed indefinitely since a nearly-resonant $\omega_l$ may lead to atomic excitation. In our calculations we considered frequencies reasonably outside such regime. The smallest detuning considered is 23.7\,nm, much larger than the detuning 1.6\,nm used in \cite{Hammes03}, finding barrier heights up to $\sim2$~mK. Our predictions could be tested experimentally by measuring the barrier height via reflection of a Bose-Einstein condensate, as done in \cite{Stehle11}.

\subsection{Two-laser force}\label{SecTwoLasers}

The following numerical investigations refer to an enriched scenario in which two laser beams shine on the back of the gold layer at the same time.
In this case the total potential is described by Eq.~\eqref{EqTwoLaserGeneral2}.

\subsubsection{Blue and red shifted lasers}

Let us begin by the case of a blue-shifted laser plus a red-shifted one. In our numeric calculations we choose $\omega_l^{\rm b}=24.6\times10^{14}$~rad/s and $\omega_l^{\rm r}=21.0\times10^{14}$~rad/s, leaving all the other system parameters as before. For these frequencies, the rubidium atom cannot follow the time-dependent oscillations of Eq.~\eqref{EqTwoLaserGeneral2}, so that their contribution averages to zero.

\begin{figure}[t]
\includegraphics[width=0.47\textwidth]{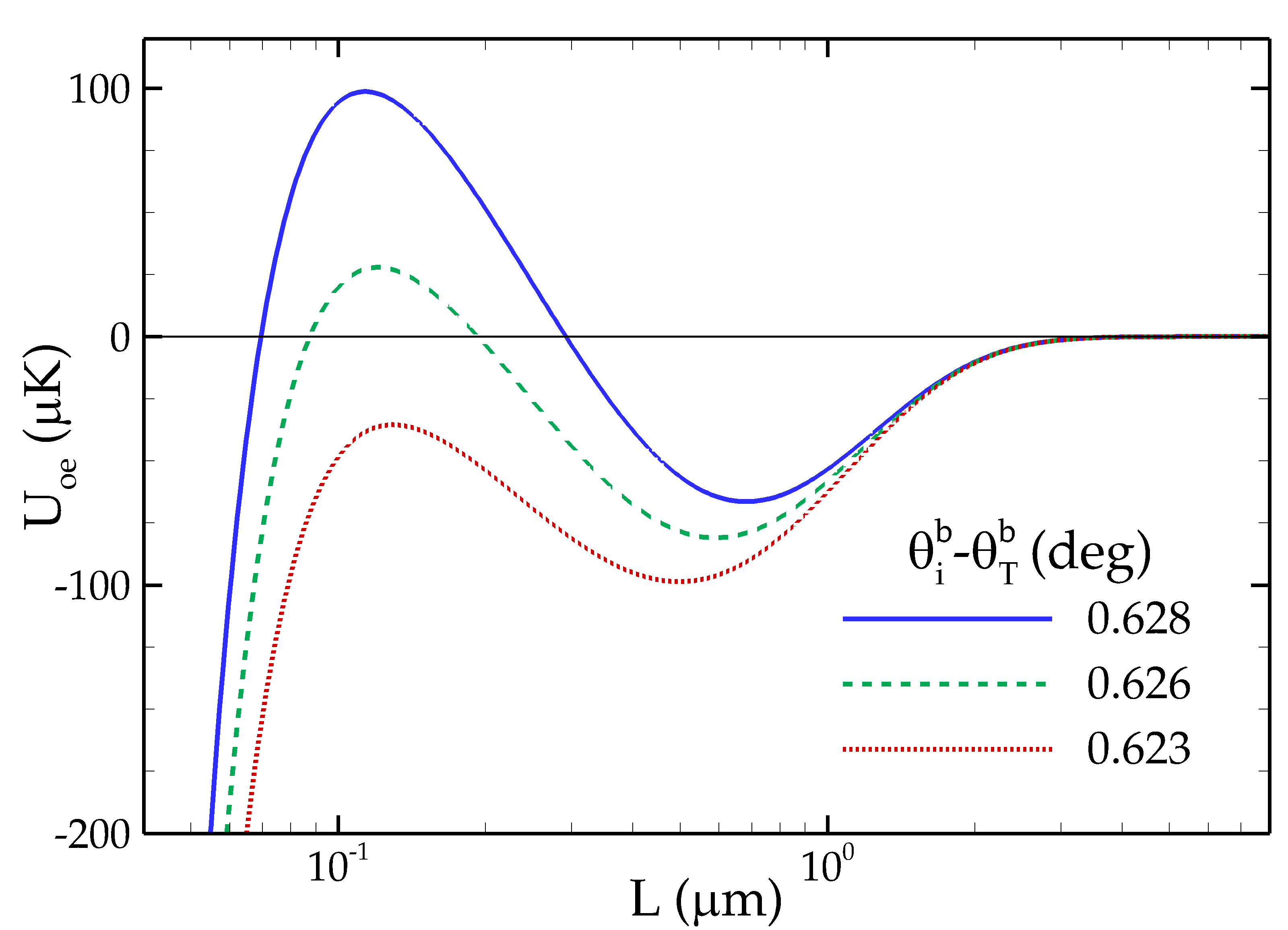}
\caption{\label{FigWells} Total out-of-equilibrium interaction energy $U_{\rm oe}$ for a rubidium atom at a distance $L$ from a structure like that of Fig.~\ref{FigKretschmann}. The present plot refers to a gold surface excited by two laser beams, of frequencies $\omega_l^{\rm b}=24.6\times10^{14}$~rad/s and $\omega_l^{\rm r}=21.0\times10^{14}$~rad/s. These frequencies are respectively blue- and red-shifted with respect to the main transitions of rudibium, whose polarizability is modeled according to Eq.~\eqref{EqRbPolarizability}. The total internal reflection angles at the bare glass/vacuum interface result $\theta_{\rm T}^{\rm b}=34.23$~deg for $\omega_l^{\rm b}$ and $\theta_{\rm T}^{\rm r}=34.63$~deg for $\omega_l^{\rm r}$. We assumed $P_l=1.2$~W and $w_l=180$~$\mu$m as power and beam waist for both beams within the glass. The plotted curves correspond to different incidence angles of the blue-shifted beam $\theta_{\rm i}^{\rm b}$ (cf. legend), while we fixed $\theta_{\rm T}^{\rm r}-\theta_{\rm i}^{\rm r}\simeq0.502$~deg, corresponding to the surface-plasmon resonance condition for the red-shifted laser.}
\end{figure}

In Fig.~\ref{FigWells} we show the total potential perceived by the atom in such a two-laser configuration as a function of its distance $L$ from the metal/vacuum interface.
The interaction results from the competition between repulsive and attractive evanescent contributions, whose decay lengths depend only on the laser frequencies. Their amplitude, instead, can be tuned via the angles of incidence at the glass/metal interface. When the amplitudes are comparable, the overcome of an evanescent contribution on the other can depend on $L$, as exemplified in Fig.~\ref{FigWells}. Here, we considered the red-shifted laser to shine always at the surface-plasmon resonance, while the blue-shifted incidence angle is varied for the three curves to change the relative weight of attraction and repulsion. Besides the barrier, already observed in the one-laser configuration, one notices the appearance of a potential well. Such kind of two-laser evanescent potential has been already observed for evanescent waves outside a dielectric \cite{Hammes03}. Here, we want to stress that the depth of such a well can be resonantly enhanced exploiting surface-plasmon resonances.

\begin{figure*}[t]
\includegraphics[width=0.33\textwidth]{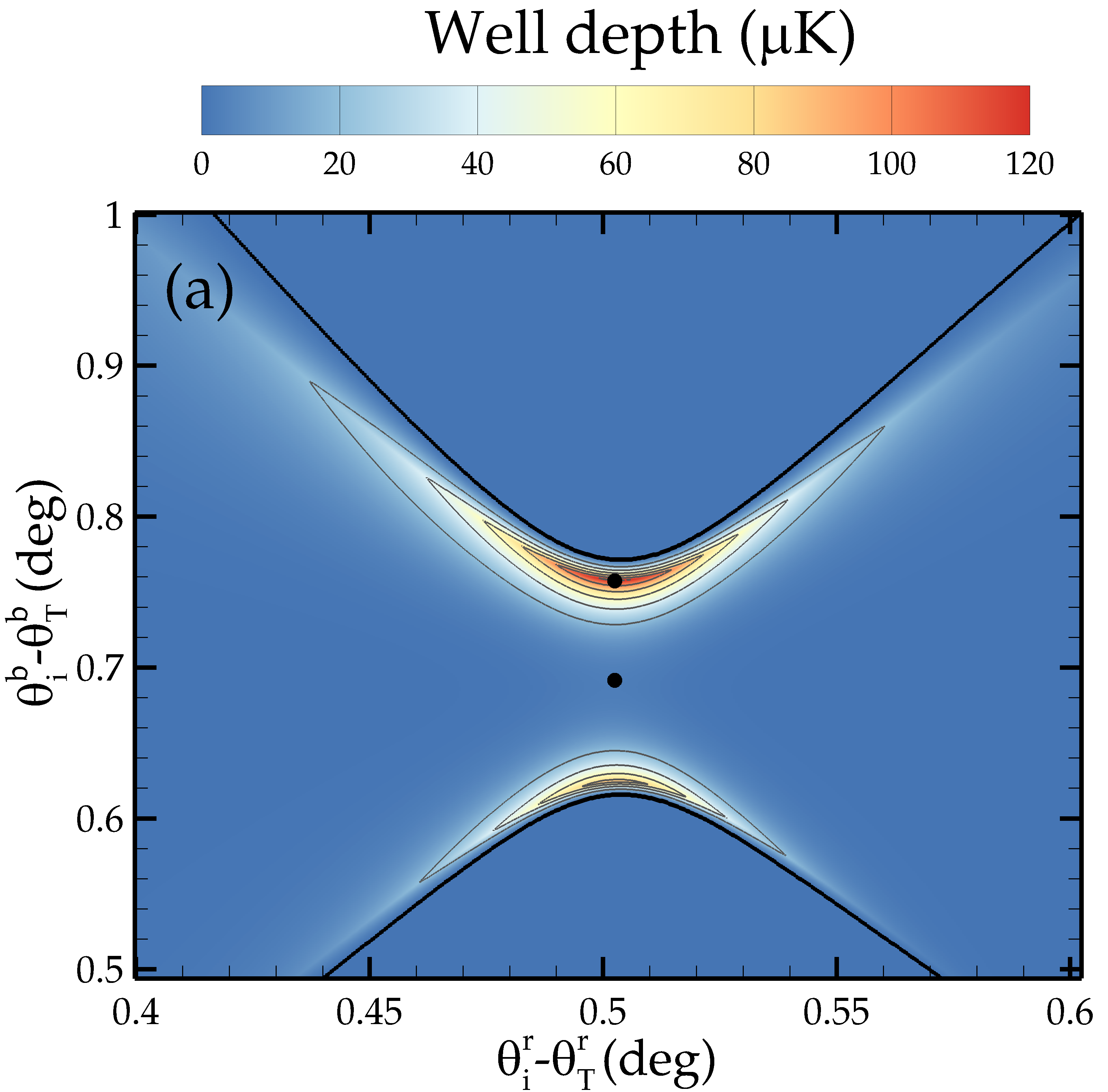}
\includegraphics[width=0.65\textwidth]{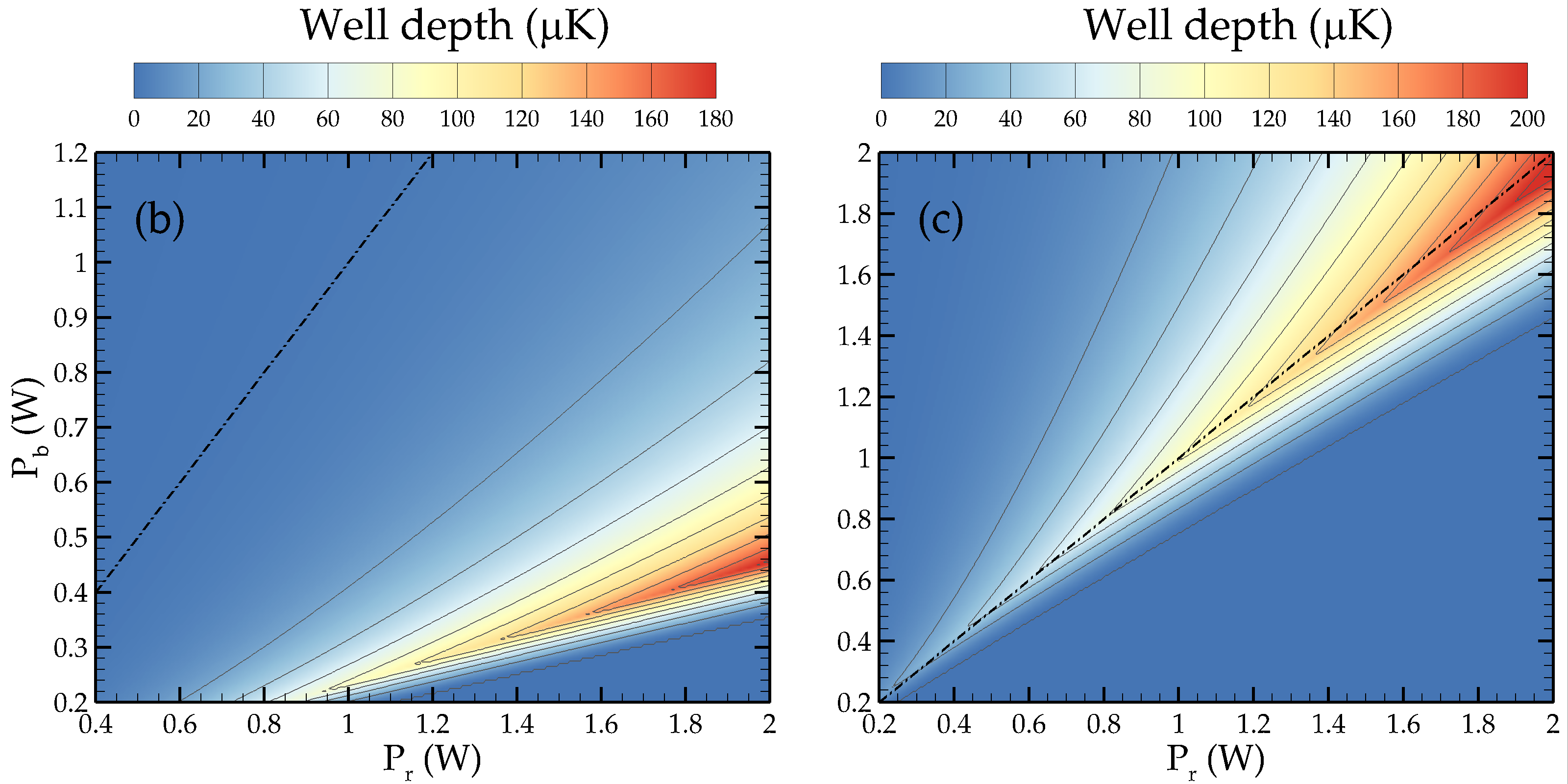}
\caption{\label{FigBoomerangPowers}
(a) Depth of the potential well generated by the total atom-surface interaction in the two-laser configuration of Fig.~\ref{FigWells}. The well depth is shown as a function of the incidence angles $\theta_{\rm i}^{\rm r}$ and $\theta_{\rm i}^{\rm b}$ of the two beams. The thick (black) contour indicates the zero-depth boundary, beyond which the well disappears. The two dots marks the incidence angles to which panels (b) and (c) refer. Panels (b) and (c)
show the depth of the out-of-equilibrium potential well as a function of the red- and blue-shifted laser powers $P_\text{r}$ and $P_\text{b}$. Panel~(b) corresponds to both lasers satisfying their surface-plasmon resonance condition, i.e. $\theta_{\rm T}^{\rm r}-\theta_{\rm i}^{\rm r}\simeq0.502$~deg and $\theta_{\rm T}^{\rm b}-\theta_{\rm i}^{\rm b}\simeq0.691$~deg. For panel~(c) we just changed $\theta_{\rm T}^{\rm b}-\theta_{\rm i}^{\rm b}\simeq0.757$~deg, correspoding to the maximum depth observed in the left panel. In both (b) and (c), the thick dot-dashed line marks $P_r=P_b$.
All other parameters are the same as in Fig.~\ref{FigWells}.}
\end{figure*}

We now analyze how the existence and the depth of the potential well are affected by the incidence angle and the power of both external lasers. We start by plotting in Fig.~\ref{FigBoomerangPowers}(a)  the depth of the potential well as a function of the incidence angles of the two beams for fixed and equal laser powers.
We observe that a potential well exists only in a given region of the diagram (between the two solid lines). In this region, the well depth is very sensitive to the incidence angles, reaching $\sim$120~$\mu$K, that is 10 times larger than typical values observed for dielectrics in similar experimental setups \cite{Hammes03}. More interestingly, one finds that the deepest well does not occur for both lasers shining at the corresponding resonant angle (cf. dot at the center of the plot). This is because $\omega_l^{\rm b}$ is much closer to the rubidium frequencies $\omega_i$ with respect to $\omega_l^{\rm r}$, which implies $|\alpha_{\rm Rb}(\omega_l^{\rm b})|>|\alpha_{\rm Rb}(\omega_l^{\rm r})|$. Hence, in order to have comparable amplitudes of the evanescent waves, one has to amplify more the red-shifted contribution than the blue-shifted one.
This picture is confirmed by the results in panels~(b) and~(c) of Fig.~\ref{FigBoomerangPowers}, where we investigate the well depth for fixed angles of incidence and varying the laser powers, respectively $P_r$ and $P_b$ for the red- and blue-shifted laser. In Fig.~\ref{FigBoomerangPowers}(b) both beams are resonant (corresponding to the dot in the center of Fig.~\ref{FigBoomerangPowers}(a)), and the deepest wells appear for $P_b<P_r$, so that the power inbalance restores the evanescent amplitudes to comparable values. Taking the angles which realize the deepest well of Fig.~\ref{FigBoomerangPowers}(a) (upper dot in the figure), one gets the behavior shown in panel~(c). This time, the deepest well appears along the $P_b=P_r$ line. Furthermore, we see that the maximum depth linearly grows with power. Such deep wells may be exploited as trapping potential for atoms in the vicinity of the metal/vacuum interface.

\subsubsection{Equal and close-frequency lasers}\label{Twolasers}

\begin{figure}[t]
\includegraphics[width=0.47\textwidth]{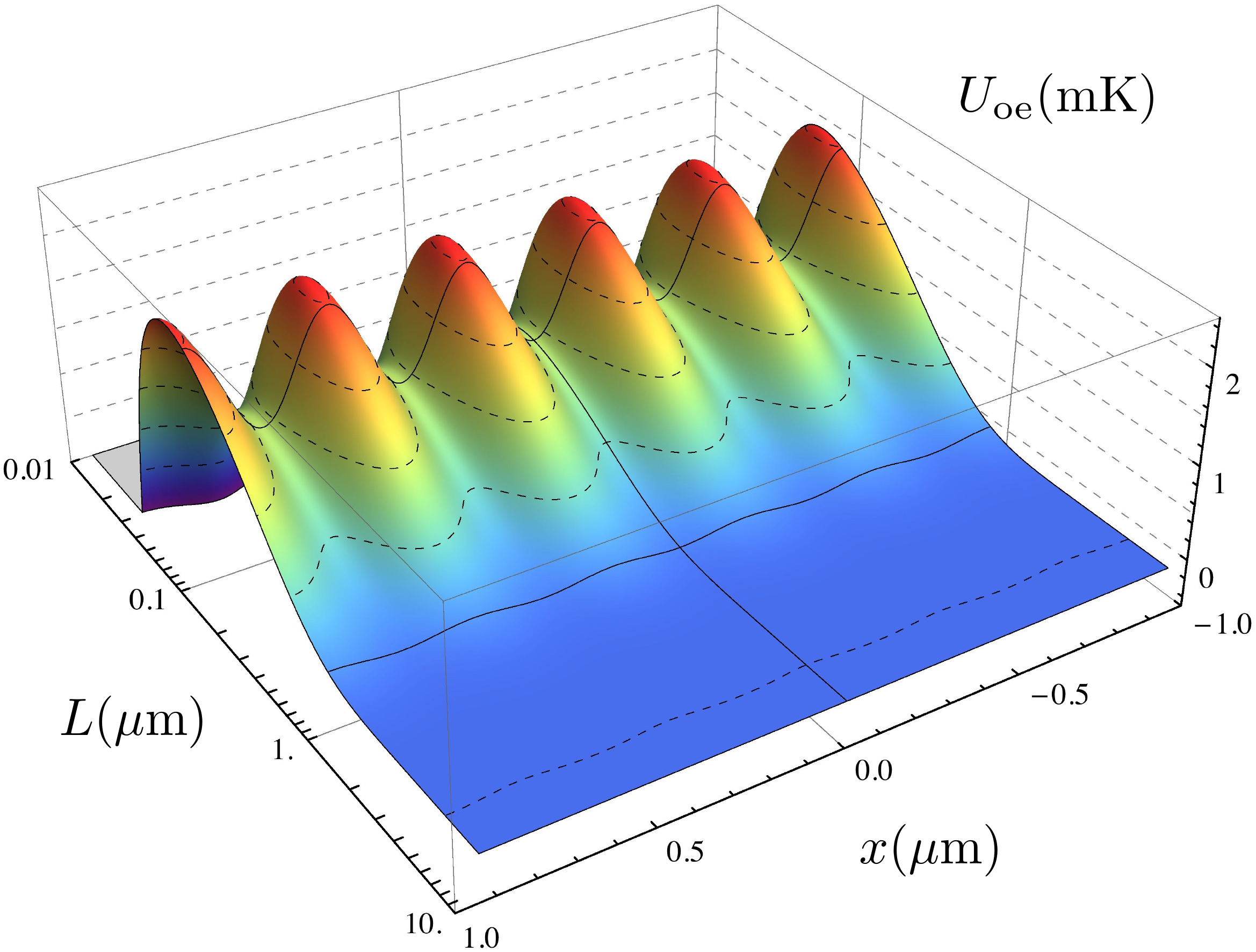}
\caption{\label{FigInterference}Out-of-equilibrium atom-surface potential $U_{\rm oe}$ for a rubidium atom at distance $L$ from a structure like that of Fig.~\ref{FigKretschmann} in the presence of two laser beams of frequency $\omega_l=24.6\times10^{14}$~rad/s, power $P_l=200$~mW, and waist $w_l=180$~$\mu$m. We consider the beams as counter-propagating along the $x$ axis, which implies that $U$ depends on $x$ but not on $y$. All the physical parameters of the system are the same of Fig.~\ref{FigFlame}.}
\end{figure}

To conclude, we present our numerical results for the case of two identical blue-shifted laser beams, propagating with opposite components of the in-plane wave vector ${\bf k}$ (cf. Fig.~\ref{FigKretschmann}), with ${\bf k}$ lying along the $x$ axis.
In this configuration, the time dependence in Eq.~\eqref{EqTwoLaserGeneral2} disappears and a standing \emph{plasmonic lattice} forms at the metal-vacuum interface. Thus, in this case Eq.~\eqref{EqTwoLaserCounter} describes the potential energy induced by the two lasers. In Fig.~\ref{FigInterference} we plot the total atom-surface potential as a function of the atom-surface distance $L$ and the atomic coordinate $x$. For the present geometry, the interaction stays $y$-independent. The total potential is now modulated along the $x$ direction, with a period $\pi/k\simeq376$~nm, that is much smaller that the typical beam waist of a laser $\sim200$~$\mu$m. The potential barrier is reduced or amplified depending on whether the counter-propagating plasmons interfere destructively or constructively. This creates a steady space-dependent potential like that obtained by replacing the uniform metallic layer by a grating \cite{Stehle11,BenderPRX14}. Similarly, the space modulation of the atom/surface interaction could be detected in the interference pattern of a back-scattered Bose-Einstein condensate.

The use of plasmonic lattices seems indeed efficient to get a spatial-dependent potential even though the metallic layer is uniform. Furthermore, one can realize more complex interference patterns resulting in structured attractive/repulsive potentials, with techniques similar to those used in the realization of optical lattices \cite{PetsasPRA94,BlochNatPhys05}. One may also switch from one potential to another without changing the glass/metal structure. Finally, we mention the possible use of two slightly-detuned laser beams to get a quasi-steady potential like that of Fig.~\ref{FigInterference}, which drifts in time slowly enough to be perceived by an atom.
Hence, the atom-surface potential would also result locally periodic in time.
This configuration could be used to transfer lateral momentum to a single particle or to stir a cloud of atoms. It could be also useful in the realization of contact-free rack and pinion systems like those proposed in \cite{AshourvanPRL07} or to induce periodic potentials near the surface without recurring to nanostructuring. A similar kind of slowly-varying interference pattern is typically used to realize the so-called shaken optical lattices.

\section{Conclusions}\label{SecConclusions}

We addressed the Casimir-Polder interaction between an atom and a surface, focusing on effects resulting from the combination of nonequilibrium effects and material properties. After a general discussion on the field and atom contributions to the energy, we analytically identified the part of the interaction due to the plasmonic-mode branch. We numerically investigated the effect of a thermal unbalance between this branch and the other field modes. We showed that this out-of-equilibrium scenario can qualitatively modify the atom-surface potential resulting in a nonmonotonous behavior. We then focused on the configuration in which one or two modes of the field are pushed out of equilibrium by external laser beams.  We calculated the laser-modified Casimir-Polder interaction for several experimental configurations showing a variety of potential landscapes, e.g. barriers, wells and periodic potentials. All these effects have been shown to be widely tailorable as a function of experimental parameters. Moreover, our results show that a realistic description of both the atom and the substrate is essential in order to describe both qualitatively and quantitatively the atom-surface interaction. As a matter of fact, the total interaction arises from a nontrivial interplay between the several different experimental parameters and the optical properties of the system.
Our findings pave the way to several possible developments. In particular, the behavior of the out-of-equilibrium energy can be explored for a variety of atomic species and dielectric properties of the substrate, and it can also be interesting to investigate the effects of nonclassical states of the field.

\section{Acknowledgments}
We thank the LANL LDRD program for financial support.
N.B. thanks Los Alamos National Laboratory for its hospitality
in the early stages of the project and the University of Palermo
for partial funding through the PerfEst 2010 scholarship. R.M.
thanks V. Parigi for useful discussions and the Center of
Nonlinear Studies at Los Alamos National Laboratory for
its hospitality. F.I. acknowledges financial support from the
European Union Marie Curie People program through the
Career Integration Grant No. PCIG14-GA-2013-631571 and
from the DFG through the DIP program (FO 703/2-1).

\end{document}